\documentclass[fleqn,usenatbib]{mnras}

\usepackage{newtxtext,newtxmath}

\usepackage[T1]{fontenc}

\DeclareRobustCommand{\VAN}[3]{#2}
\let\VANthebibliography\thebibliography
\def\thebibliography{\DeclareRobustCommand{\VAN}[3]{##3}\VANthebibliography}

\usepackage{float}      %
\usepackage{graphicx}   
\usepackage{amsmath}    
\usepackage{natbib}     
\usepackage{longtable} 
\usepackage{booktabs}   
\usepackage{caption}    
\usepackage{soul}     
\usepackage{ulem}     

\title[TESS photometry of massive stars in the SMC]{Binarity at LOw Metallicity (BLOeM): massive star variability revealed using a novel software tool for point-spread function fitting of TESS images\\}

\author[P. Van Daele et al.]{
Pieterjan J. Van Daele$^{1}$\thanks{E-mail: p.van-daele2@newcastle.ac.uk},
Dominic M. Bowman$^{1,2}$,
Roey Ovadia$^{3}$,
Zehava Katabi$^{3}$,
Julia Bodensteiner$^{4}$,
\newauthor
Tomer Shenar$^{3}$,
Norbert Langer$^{5}$,
Jan Henneco$^{1}$,
Ankur Kalita$^{1}$,
Paul A. Crowther$^{6}$,
Maude Gull$^{7,8}$,
\newauthor
Laurent Mahy$^{9}$,
Lee Patrick$^{10}$,
Daniel Pauli$^{2}$,
Micha{\l} Pawlak$^{11}$
\\
$^{1}$School of Mathematics, Statistics and Physics, Newcastle University, Newcastle upon Tyne, NE1 7RU, UK\\
$^{2}$Institute of Astronomy, KU Leuven, Celestijnenlaan 200D, B-3001 Leuven, Belgium \\
$^{3}$The School of Physics and Astronomy, Tel Aviv University, Tel Aviv, 6997801, Israel\\
$^{4}$Anton Pannekoek Institute for Astronomy, University of Amsterdam, Science Park 904, 1098 XH Amsterdam, The Netherlands\\
$^{5}$Argelander-Institut für Astronomie, Universität Bonn, Auf dem Hügel 71, 53121 Bonn, Germany\\
$^{6}$Astrophysics Research Cluster, Mathematical \& Physical Sciences, University of Sheffield, Hounsfield Road, Sheffield, S3 7RH, UK\\
$^{7}$The Observatories of the Carnegie Institution for Science, 813 Santa Barbara Street, Pasadena, CA 91101, USA\\
$^{8}$Department of Astronomy, California Institute of Technology, Pasadena, CA 91125, USA\\
$^{9}$Royal Observatory of Belgium, Avenue Circulaire/Ringlaan 3, B-1180 Brussels, Belgium\\
$^{10}$Centro de Astrobiología (CSIC-INTA), Ctra. Torrejón a Ajalvir km 4, 28850 Torrejón de Ardoz, Spain\\
$^{11}$Lund Observatory, Division of Astrophysics, Department of Physics, Lund University, Box 43, SE-221 00, Lund, Sweden\\
}

\date{Accepted 2026 June 09; in original form 2026 March 19}

\pubyear{\the\year{}}

\begin{document}
\label{firstpage}
\pagerange{\pageref{firstpage}--\pageref{lastpage}}
\maketitle

\begin{abstract}
Massive stars, the progenitors of neutron stars and black holes, play a crucial role in shaping the chemical and radiative properties of entire galaxies through their winds and explosive deaths. Stellar pulsations are a common phenomenon in massive stars and asteroseismology -- the study of such pulsations -- provides crucial constraints on the physics of massive star interiors. The excitation of heat-driven pulsations in massive stars is expected to depend on a star's metallicity, but this remains largely uncalibrated in evolution models due to a lack of a sufficient observations. While TESS has dramatically improved the statistics for Galactic massive stars, obtaining TESS light curves for low-metallicity massive stars beyond the Milky Way is challenging, due to their faintness and heavy crowding. In this paper, we present a novel point-spread function (PSF) based light curve extraction method called {\sc Lemons}, which overcomes these challenges. We also demonstrate the limitations of the often-used simple aperture photometry (SAP) method that can provide heavily contaminated light curves. With this new technique, accurate light curves of 91 SMC massive stars in the BLOeM sample are extracted. They reveal a variety of variability types including indications of binarity (e.g. eclipses and ellipsoidal modulation) and stellar pulsations. They also enable us to investigate stochastic low-frequency (SLF) variability for massive stars in the SMC. Furthermore we demonstrate how the morphology of SLF variability probes a star's location in the Hertzsprung--Russell diagram, which appears similar to Galactic massive stars thus indicating that the underlying physical mechanism could be insensitive to metallicity.

\end{abstract}




\section{Introduction}

Massive stars have birth masses above about 8~M$_{\odot}$ and are progenitors of supernovae events, which means they are also the precursors of neutron stars and black holes \citep{Maeder2000, Kippenhahn2012, Langer2012}. Through their radiative winds and explosive deaths, massive stars are important chemical factories in the Universe, and play a major role in the study of star formation, galactic evolution \citep{Hopkins+2014}, binary systems \citep{Sana+2012, MarchantBodensteiner2024}, and gravitational waves \citep{Abbott+2016a}. 

Current stellar evolution models, however, have large uncertainties in how these stars are born, evolve and die. Specifically, we lack accurate prescriptions of interior rotation, chemical mixing, and the transport of angular momentum (see \citealt{Aerts2019_angmomReview}). Moreover, ongoing research has shown the important role of mixing and angular momentum transport by internal gravity waves (IGWs; \citealt{Rogers2015, Rogers+2017, Rogers2025, Varghese+2023, Varghese2025}) in massive stars, which is currently not fully implemented in 1D stellar evolution models. With more efficient internal mixing, unprocessed hydrogen is transported from the envelope into the core and made available for nuclear burning, which allows the star to spend a longer amount of time on the hydrogen-burning main sequence, resulting in a larger helium core mass at the end of the main sequence (see \citealt{Bowman_2020_review, Johnston2021}). In turn this affects the evolution beyond the main sequence and therefore also the supernovae chemical yields \citep{Maeder2000, Kippenhahn2012, Langer2012}.

Additionally, most massive stars are born in binaries and multiple systems \citep{Sana+2012}, and even apparently single massive stars may be binary interaction products \citep{Langer2012, MarchantBodensteiner2024}. Therefore, a full picture of stellar evolution also requires an understanding of the effects of binarity, such as mass and angular momentum transfer, accretion, as well as stellar merger (see e.g. \citealt{Renzo+2021,Richards+2025,Henneco+2024b}). In this paper, we use ‘binarity’ as a general term to cover all possible manifestations of geometrical aspects (e.g. eclipses), physical processes (e.g. tides), and evolutionary outcomes expected from binary or multiple star systems (e.g. mergers) --- see \citet{DeMarco+2017} and \citet{Schneider2025} for reviews. These effects impact the stars internal structure and can thus alter the nature of stellar oscillations \citep{Miszuda+2021,Wagg_2024}.

The ongoing advances in asteroseismology, the study of stellar pulsations \citep{Aerts2010_book}, enables us to empirically probe and constrain the physical mechanisms and processes inside stars (see \citealt{Aerts2021_review}). Historically, coherent heat-driven pulsations have been exploited for massive stars for several decades (see \citealt{Bowman_2020_review} for a review), which are excited by an iron-opacity bump near the stellar surface for masses above about 3\,M$_{\odot}$ and ages that span the entire main-sequence and beyond \citep{DziembowskiPamyatnykh1993, Miglio+2007, Burssens+2020}. More recently, damped pulsation modes (i.e. propagating waves rather than coherent standing waves) generated at the turbulent interface of convective and radiative regions have been detected for hundreds of massive stars in the Milky Way galaxy \citep{ Bowman+2019_nat, Bowman+2019_corot, Bowman+2020aa}, but also for massive stars in the LMC and SMC galaxies \citep{Bowman+2019_nat, BowmanVanDaele2024}. This new type of asteroseismic signal is referred to as stochastic low-frequency (SLF) variability, and has great potential for probing the interior physics of massive stars that lack coherent heat-driven pulsation modes (see \citealt{Bowman2023} for a review). This signal is thought to be caused by IGWs excited at the boundary of convective regions based on hydrodynamical simulations of stellar interiors \citep{Rogers2013, AertsRogers2015, Rogers+2017, Edelmann+2019ApJ, Schultz2022a, Anders+2023, Thompson+2024_3Dsim}.

As the photometric manifestation of IGWs, the observed SLF variability of a massive star signifies the dynamical processes occurring in its photosphere, which presumably should also introduce a spectroscopic signal. For example, for the well-studied bright galactic O-type star $\zeta\,$Oph, it has indeed been shown that the stochastic line broadening known as macroturbulence \citep{SimonDiaz+2017} contains the same pulsation periods as its photometric signal, which demonstrates a physical connection between macroturbulence and both stochastic and coherent pulsations \citep{Kalita+2025}. The corresponding velocity fluctuations seen in spectroscopy, which occur in the vicinity of the sonic point of the stellar wind, are thought to introduce or enhance so-called wind clumping \citep{Cantiello+2009,Debnath+2024}, which may strongly affect the star's mass and angular momentum loss \citep{Langer1998,Muijres+2011}. Therefore, understanding SLF variability as the manifestation of IGWs in massive stars has widespread implications for stellar structure and evolution theory.

Stars in the SMC have a lower metallicity compared to Galactic stars (i.e. $Z_{\rm SMC} \simeq 0.2\,Z_{\odot}$; \citealt{Hunter+2007}). Therefore, the opacity in the sub-surface convection zones of massive stars is lower compared to Galactic stars. This means that convection is less efficient and may carry only a small fraction of the flux, or sub-surface convection zones may even be absent entirely. For example, even for Galactic massive stars, convection is only a minor contributor to the total flux and energy transport in these turbulent sub-surface regions (see e.g. \citealt{Debnath+2024}). Since low-metallicity main-sequence massive stars may lack sub-surface convection zones entirely \citep{Jermyn+2022, BowmanVanDaele2024}, SLF variability in such stars would more likely originate from the only remaining convection zone: the convective core. \citet{BowmanVanDaele2024} performed a proof-of-concept study of studying SLF variability in LMC and SMC massive stars and found little difference in the amplitudes or dominant frequency range of SLF variability compared to Galactic massive stars. However, a systematic and detailed characterisation of SLF variability for SMC massive stars across the Hertzsprung--Russell (HR) diagram is yet to be performed. This is important to constrain, particularly because IGWs are efficient transporters of chemical species and angular momentum (e.g. \citealt{Rogers+2017, Rogers2025, Varghese2025}). Hence, IGWs and SLF variability represent crucial pieces of the stellar evolution puzzle.

In addition to different types of stellar pulsations, variability in a light curve can also have other causes such as binarity or chemical spots on a rotating (chemically peculiar and/or magnetic) star, for which photometric light curves complement spectroscopic studies. In this paper we make use of a large sample of massive stars in the SMC from the BLOeM consortium (\citealt{BLOeM2024}; see also Section~\ref{sec: sample selection}). BLOeM (Binarity at LOw Metallicity) is a survey that assembled multi-epoch spectroscopy of 925 SMC massive stars and studies various aspects of massive stars, but has a focus on binarity and multiplicity. Recent work includes the estimation of the binary fraction of blue supergiants in the SMC \citep{Britavskiy+2025}, the multiplicity properties of BAF-type supergiants \citep{Patrick+2025} and rapidly rotating Oe and Be stars \citep{Bodensteiner+2025}, and the empirical definition of the terminal-age main sequence (TAMS) based on rotation rates \citep{Lennon+2026}. Furthermore, \citet{Sana+2025NatAs} showed that a high proportion of O‑type stars in the SMC have a close companion, which implies that binary interaction likely plays a dominant role in shaping massive star evolution in metal-poor environments, and by extension in the early Universe. Obtaining light curves for BLOeM targets is therefore not only necessary for the study of stellar pulsations and SLF variability for massive stars in the SMC, but also provides a valuable complementary data set for the BLOeM project to complement spectroscopic studies.

In this paper, we describe a novel software tool to extract high-quality light curves from the NASA Transiting Exoplanet Survey Satellite (TESS; \citealt{Ricker2015, TESS_2021}) mission, with a specific focus on massive stars in the SMC, which is described in detail in Section~\ref{sec: TESS}. However, this software is broadly applicable to all variable stars in crowded areas (i.e. clusters). In this work, we extract light curves of a subsample of the BLOeM targets that allow for the classification of different types of variability, such as coherent pulsations, SLF variability, stars with rotational modulation caused by spots, and binarity, which we describe in Section~\ref{sec: bloem phot}. In Section~\ref{sec:SLF}, we study SLF variability in these low-metallicity massive stars of the BLOeM sample, and we conclude in Section~\ref{sec: conclusions}.

\section{TESS light curve extraction}
\label{sec: TESS}

\subsection{Sample selection}\label{sec: sample selection}

\begin{figure*}
    \centering
    \includegraphics[width=\linewidth]{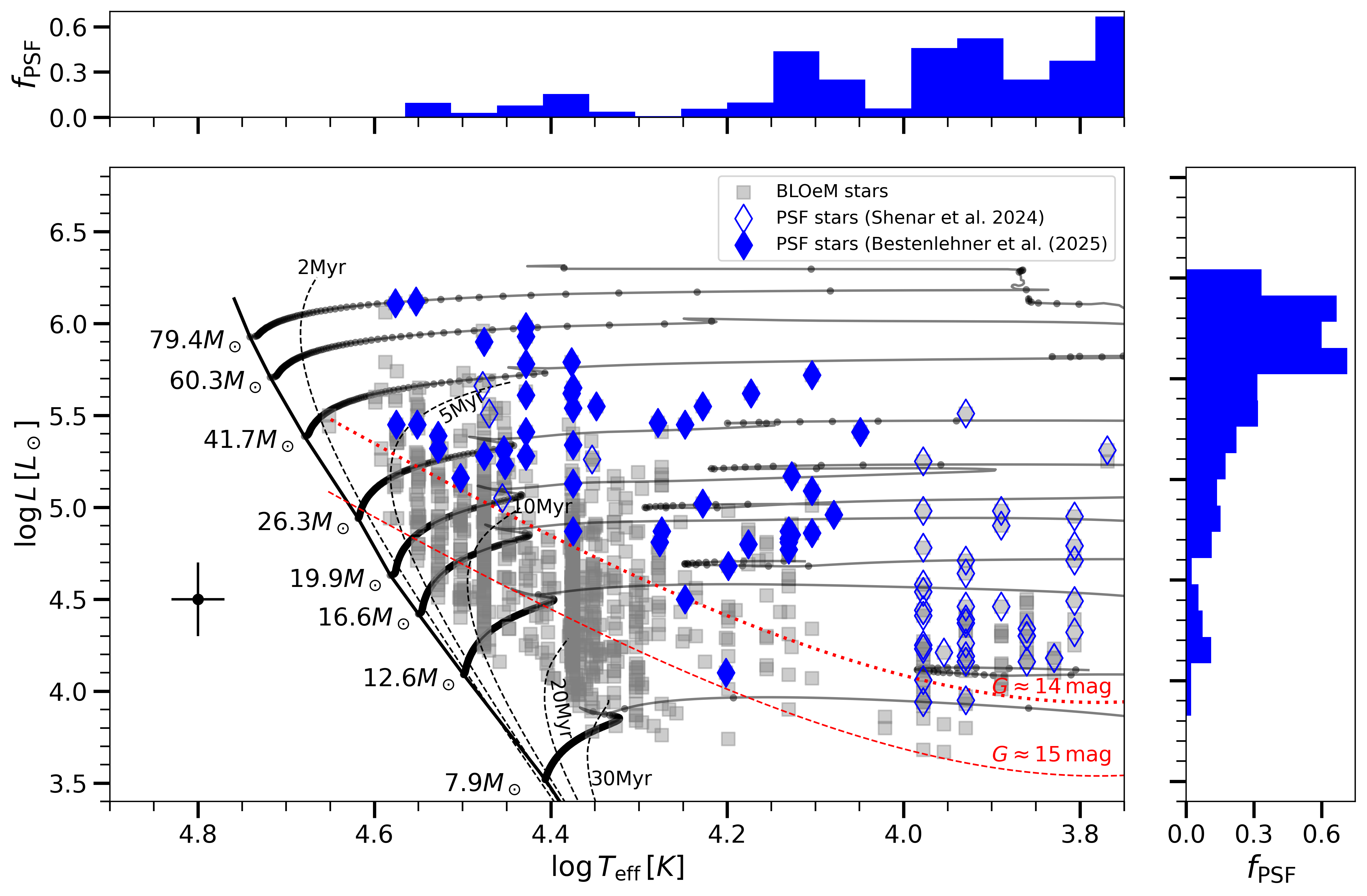}
    \caption{The locations in the HR~diagram of all BLOeM targets \citep{BLOeM2024} are shown in grey using parameters from \citet{Bestenlehner+2025} and \citet{BLOeM2024}. The targets with successfully extracted PSF TESS light curves are overplotted as blue diamonds and are delimited by the TESS brightness limit of about $G < 15$~mag, which is indicated by the red-dashed line. Note that PSF sub-sample is biased towards brighter targets and thus does not include the full diversity of the BLOeM population. The histograms along each axis show the completeness fraction (i.e. the ratio of targets with PSF light curves relative to the number of total BLOeM stars in that bin for each axis). A typical error bar for the BLOeM spectroscopic parameters is shown in the bottom-left, as well as labelled evolutionary tracks and isochrones from \citet{Schootemeijer+2019}, with dots on the evolutionary tracks marking equal time steps of 0.05~Myr.}
    \label{fig:HRD BLOeM}
\end{figure*}

\begin{figure}
    \centering
    \includegraphics[width=\linewidth]{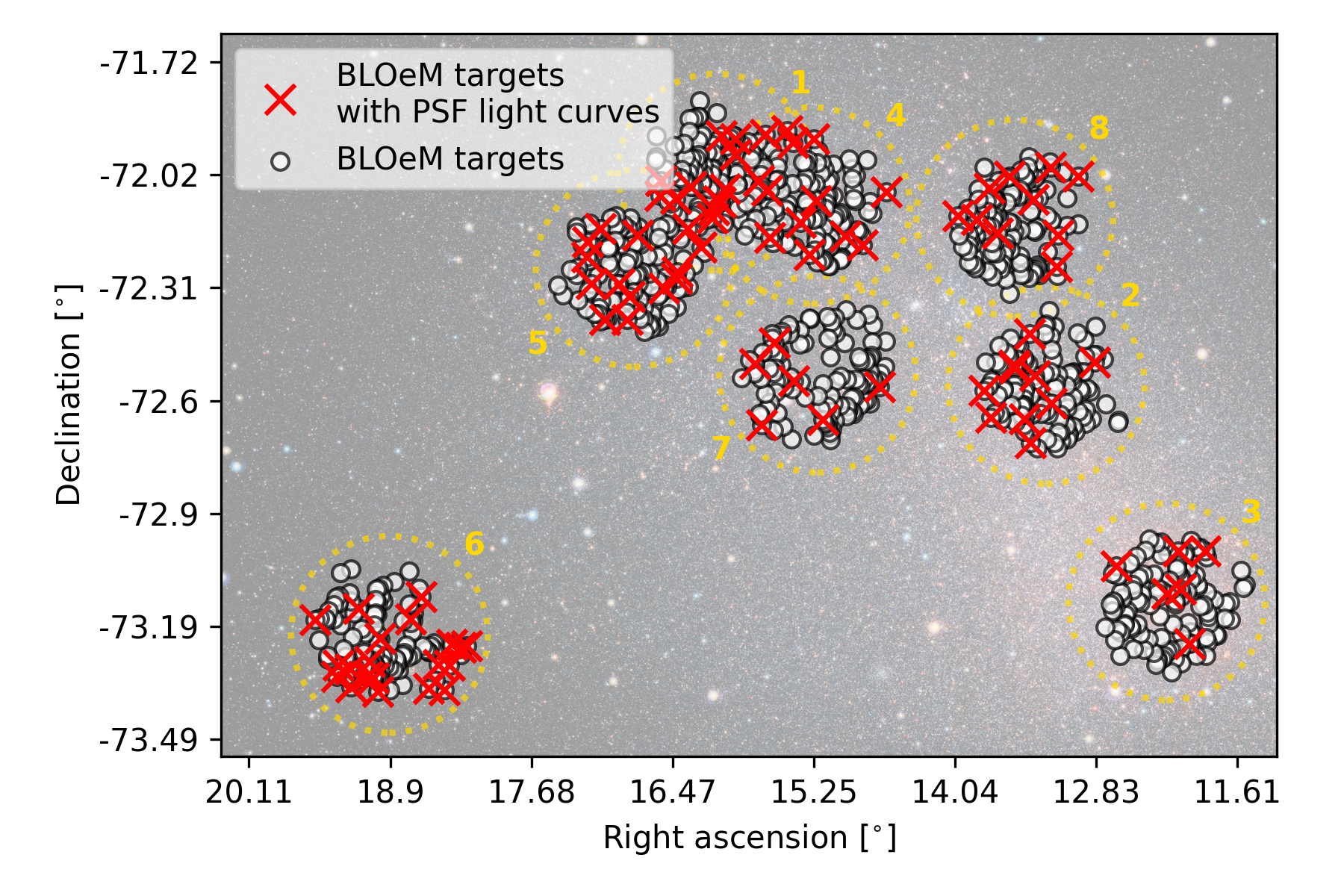}
    \caption{Spatial location of the BLOeM targets for which a successful PSF light curve have been extracted. Figure adapted from \citet{BLOeM2024}.}
    \label{fig:ra&dec BLOeM}
\end{figure}

The BLOeM campaign provides a sample of 925 massive stars in the SMC \citep{BLOeM2024}, which are interesting objects for asteroseismology and are excellent test cases for our new PSF photometry method (see Section~\ref{section: lemons}). This sample consists of main sequence (OB-type) and evolved (OBAF-type) massive stars, with masses ranging from about $7$ to $60$~M$_\odot$ and {\sc Gaia} $G$-band magnitudes ranging from 10.1 down to 16.4 in the SMC. The HR~diagram of the full BLOeM sample is presented in Fig.~\ref{fig:HRD BLOeM} using spectroscopic parameters from \citet{BLOeM2024} and when available the revised parameters determined by \citet{Bestenlehner+2025}. The location of the BLOeM sample on the sky is shown in Fig.~\ref{fig:ra&dec BLOeM}, which also indicates the stars with new PSF light curves extracted in this work (see Section~\ref{sec: bloem phot}).

\subsection{The TESS mission and data}
\label{sec: TESS mission}

In this work we use time-series photometry assembled by the TESS space mission \citep{Ricker2015,TESSmission_2021}. TESS is a full-sky survey launched in 2018 with a primary goal to detect the transits of exoplanets orbiting bright stars in our Galaxy, but these data also allow for high-precision asteroseismic studies of massive stars (see e.g. \citealt{Bowman+2019_nat, Burssens+2020, Burssens+2023}). TESS is currently in its second extended mission and has to-date delivered light curves spanning several years for millions of stars across the sky. With particular attention to extracting SMC light curves with minimal contamination, the limiting magnitude is about 15~mag for TESS, which corresponds roughly to stars with initial masses above about 10~M$_{\odot}$ on the main sequence in the SMC (see \citealt{BowmanVanDaele2024}). This approximate brightness limit of TESS in the HR~diagram is shown as a red-dashed line in Fig.~\ref{fig:HRD BLOeM}. Therefore, due to the specifications of the TESS mission, extracting reliable photometry is only possible for sufficiently bright and isolated BLOeM targets.

The reduced full-frame image (FFI) data from the TESS mission are publicly available \citep{Jenkins+2016, TESS_2021} from the MAST archive\footnote{\url{https://archive.stsci.edu/missions-and-data/tess}}. With the {\sc Lightkurve} \citep{Lightkurve} and {\sc TESScut} \citep{TESScut_ref} \texttt{python} packages, we searched for a target's coordinates, and downloaded a cut-out of the FFI centered on a target. Typically, cut-outs of at least 19 $\times$ 19 pixels\footnote{An odd number was chosen to conveniently define fixed SAP masks around the central pixel.} are needed to capture all the target signal as well as sufficient background pixels for background subtraction as part of the reduction procedure.

Simple aperture photometry (SAP) is the standard approach to extract a light curve from TESS FFI data. This is done with a so-called binary aperture pixel mask, such that pixels are assigned to either capture the target flux or not. The total flux is then the sum of the flux of all pixels in the selected aperture mask for each time stamp. The aperture mask is typically defined as consisting of pixels that contain flux above a certain threshold chosen based on the level of contamination and background around the target. See, for example, \citet{Cubespec_Bowman2022} and \citet{Stefano_2022} for applications of SAP to TESS data for early-type stars. This is a sufficiently good approach for bright, well-isolated Galactic stars (see discussion by \citealt{Scott+2026}), but starts to fail for fainter stars in more crowded regions. This is because the light of multiple nearby stars blends together in crowded regions, such as the SMC, meaning it becomes non-trivial to define an aperture mask without large amounts of contamination. Moreover, it becomes challenging to identify which signal (e.g. pulsation mode frequency or eclipses) originates from which star. In a minority of cases, one could tackle the issue of contamination by only looking at the light curve of a single pixel in which the flux of the target of interest may dominate (e.g. \citealt{Higgins2023, Pauli2023a, Fritzewski+2025}), but this is more susceptible to unwanted pixel-to-pixel sensitivity and instrumental trends.

\subsection{Light curve extraction with {\sc Lemons}}\label{section: lemons}

\begin{figure}
    \centering
    \includegraphics[width=\columnwidth]{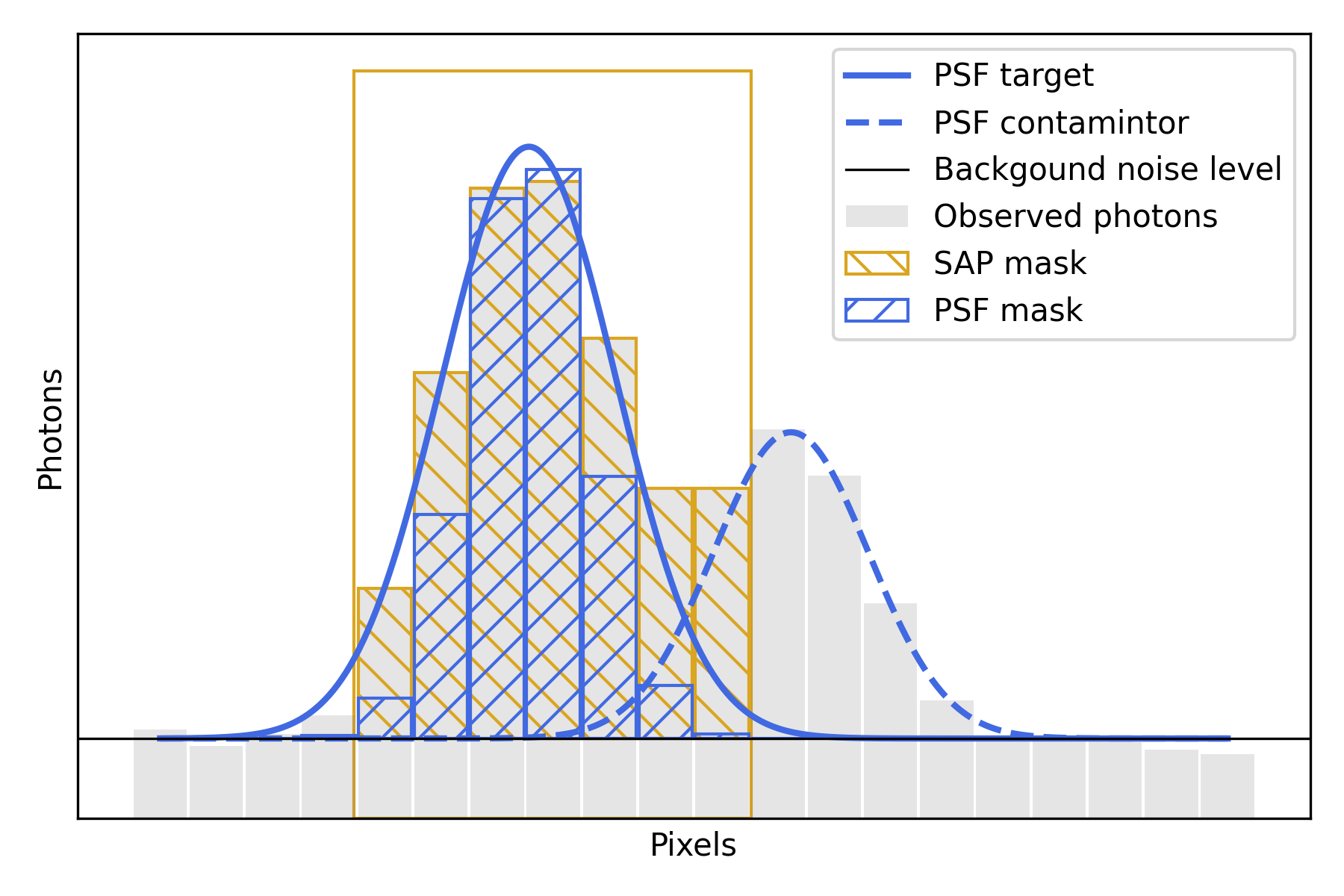}
    \caption{A 1-dimensional schematic of PSF fitting on top of a background flux, applied on two artificial targets that blend together. The golden hatched area represents a typical SAP mask that captures unwanted contaminated signal from a nearby source, but is minimised in a PSF fitting approach.}
    \label{fig:psf_schematic}
\end{figure}

In this work, we have developed a novel light curve extraction software package called {\sc Lemons}\footnote{\url{https://github.com/pieterjanv314/Lemons}} that models how the light of a point source is distributed over a TESS FFI image --- the point spread function (PSF). Our approach achieves a preferable balance between minimised flux contamination while incorporating as much astrophysical signal from the target star as possible in crowded regions. Our PSF approach is more robust against contamination compared to the commonly used SAP methodology because the pixels closer to a star's location on the CCD are prioritised when using a PSF approach, thus they contribute a higher weighted fraction in extracting a target's light curve. We show a 1D schematic of a Gaussian-like PSF in Fig.~\ref{fig:psf_schematic}, which demonstrates its advantages to the SAP approach for close sources.

PSF fitting of TESS FFI images to extract light curves is certainly not a new idea and has been done for studies targeting low-mass stars in the galaxy. For example, existing packages include {\sc Eleonor} \citep{Eleonor_2019}, {\sc Pathos} \citep{PATHOS}, and {\sc TGLC} \citep{Han+2023_tglc}. However, these tools are typically optimised for exoplanet searches around low-mass stars and therefore often include light curve post processing techniques that potentially modify stellar variability. After initial testing of available software tools applied to extra-Galactic massive stars (see \citealt{BowmanVanDaele2024}), we have developed a new software tool optimised for studying massive star variability. Our approach has the advantages of including centroid tracking informed by ultra-precise co-ordinates assembled by ESA's {\sc Gaia} mission \citep{Gaia2016, Gaia2021, Gaia2023}, as well as accounting for variations in the size of the PSF per image frame, which is necessary for variable stars with large peak-to-peak variability (e.g. eclipsing binaries and high-amplitude pulsators).

\subsubsection{Implementation of point spread function fitting}

The {\sc Lemons} code developed in this work makes use of the {\sc psf} module within the {\sc photutils} package \citep{photutils}, which contains tools for PSF photometry. It performs a least-squares fit of 2D Gaussian PSFs, on top of an estimated background level with a Levenberg-Marquardt algorithm. This algorithm optimises parameters $\beta$, consisting of a height (amplitude), position (i.e. centroid) and broadness parameter (i.e. standard deviation of the 2D Gaussian), to minimise the $\chi^2$ cost function 
    \begin{equation}
        \chi^2(\beta) = \sum_{i}\left[ f_i - (G_i(\beta) +b)\right]^2
    \end{equation}
where $i$ enumerates all pixels in the image, $f_i$ and $G_i(\beta)$ represent the measured flux and the one predicted from the Gaussian PSF, respectively. Allowing centroid and broadness to be free parameters is key to our optimisation for variable stars. The background flux $b$ is estimated as the average flux of all pixels fainter than 1~per~cent of the median flux of the image excluding pixels containing sources. Relatively large TESS FFI cutouts of $19 \times 19$ pixels are used to ensure that sufficient background pixels are included. The final flux of a star in an image is extracted as the weighted sum over the pixel flux value, where different weights are assigned to each pixel according to the Gaussian PSF shape integrated over the pixel's area.

In our initial testing, we found that differences in the optical path for stars at different locations in the CCDs across all four TESS cameras as well as variable scattered light mean that the PSF of a target is both time and spatially dependent \citep{VanDaele_MScThesis} --- see also discussion by \citet{Han+2023_tglc}. 
However, the tests performed by \citet{VanDaele_MScThesis} illustrated that the assumption of circularly symmetric PSFs are sufficient for studying massive star variability given the relatively large pixel size (21~arcsec) of TESS's cameras. In the case of variable scattered light, as long as the star is not too close to the edge of a CCD such that sufficient nearby background pixels are available, circularly symmetric PSFs provide reasonable results. {\sc Lemons} has a built in function to check and avoid cases where the target is too close to the CCD edge.

\subsubsection{Optimisation for stellar variability}
\label{sec:PSF for variable stars}

Our approach is optimised to study stellar variability in two main ways. Each represents a significant advantage to other open-source PSF-based light curve extraction tools currently available. 

Firstly, the broadness (and amplitude) of the PSF is allowed to vary over time, which means that the full-width half-maximum (FWHM) of the Gaussian PSF is a free parameter. This is necessary because as an object becomes brighter, the light tends to spread out more on the detector. This is known as the `brighter fatter' effect and occurs when a pixel receives enough electrons (due to the large number of incoming photons on the detector) that it influences the electric field locally, and causes electrons to spill over to neighbouring pixels \citep{Howell_S_2006a}. Since massive stars are often variable in their light curves (see \citealt{Bowman_2020_review}), sometimes with peak-to-peak brightness variations as large as 1~mag on times scales of hours-to-weeks, the shape of the PSF must be allowed to vary in time.

\begin{figure}
    \centering
    \includegraphics[width=\linewidth]{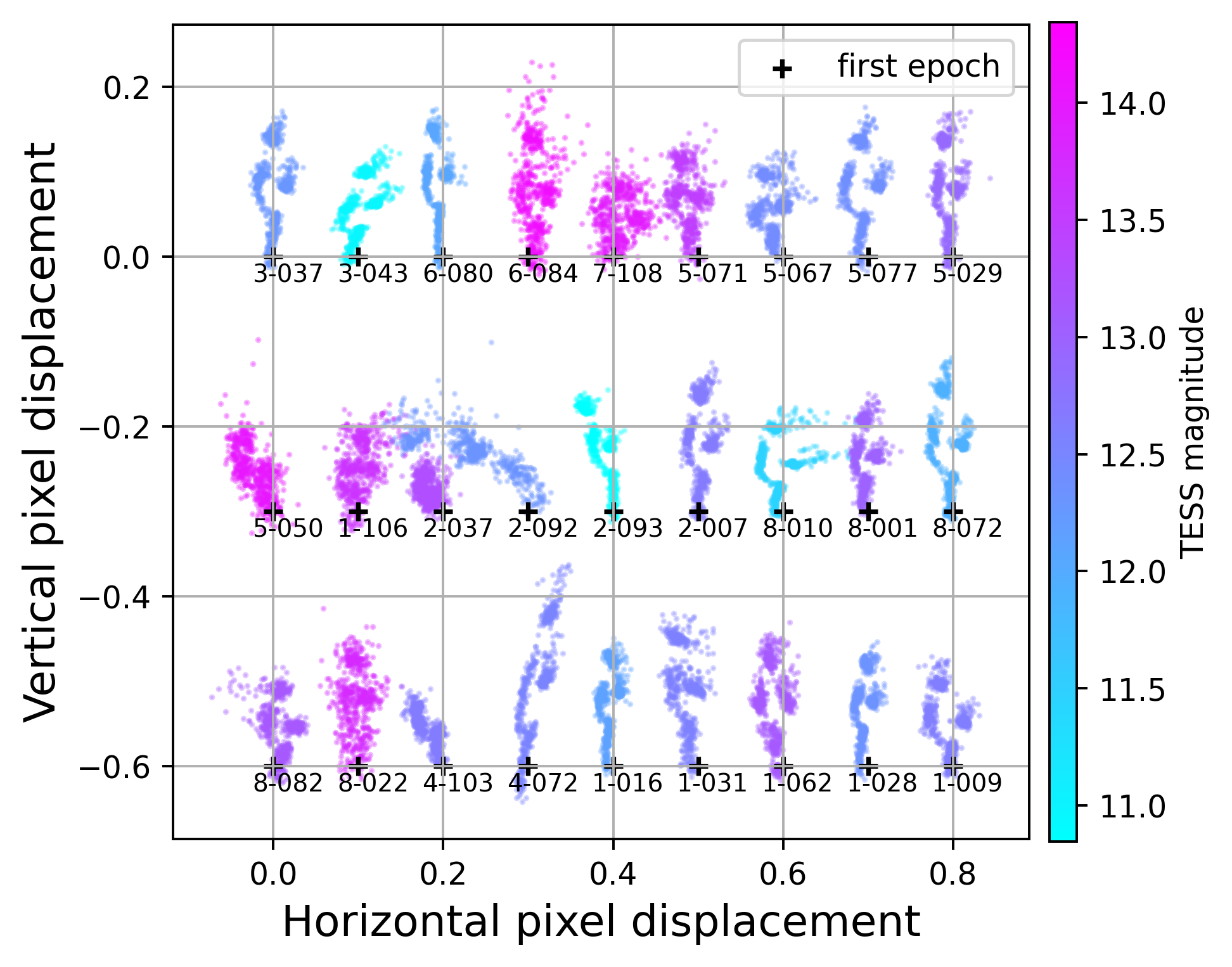}
    \caption{The relative path of the centroid for sector 68 of selected SMC massive stars, each labelled with their BLOeM identifier and colour-coded with the stars TESS magnitude. The 1$\sigma$ errors on these centroid positions are typically $0.05\sim0.1$ in terms of the pixel size.}
    \label{fig:centroid}
\end{figure}

Secondly, the centre of light, which represents the position of the star on the image, is also a free parameter for all TESS images. The {\sc Gaia} space mission \citep{Gaia2016,Gaia2021,Gaia2023} has measured the precise locations of billions of stars, and we use these positions to precisely identify stars (i.e. BLOeM targets as well as contaminating stars) in TESS FFI images. However, the pointing of TESS and its world coordinate system is not as precise as {\sc Gaia}'s, hence it is a significant advantage to be guided by {\sc Gaia} and allow the centroid to be a free parameter. Typically, the fitted PSF centroids in TESS FFIs deviate by about a tenth of TESS's 21-arcsec pixel size (corresponding to a few arcseconds on the sky) from the input {\sc Gaia} position. Additionally, there is an apparent movement of the PSF during each TESS sector because of drift and roll of the spacecraft, which is demonstrated in Fig.~\ref{fig:centroid} for a representative subset of 27 SMC massive stars. The discontinuities in groups of points in Fig.~\ref{fig:centroid} correspond to the data downloading gap that occurs near the middle of each TESS sector, in which the spacecraft points back towards Earth and then is re-pointed back on target. Naturally, this results in a different systematic pointing error both before and after the mid-sector data downlink. Furthermore, Fig.~\ref{fig:centroid} illustrates that the PSF fitting of fainter stars is generally less precise in terms of centroid grouping, and that stars in the same sector typically have similar x- and y-axis displacement patterns, which demonstrates how this effect is mainly instrumental in origin. 

Importantly, our tests revealed that it is not advisable to assume a fixed location for a PSF using the {\sc Gaia} location nor a target's average location, because flux enters and leaves a fixed-centroid Gaussian over time due to instrumental drift and therefore could introduce slow trends the extracted light curve. This is also true for the SAP approach when only a single pixel is used. Therefore, including the PSF centroid as a free parameter in all fits to the TESS image data, whilst providing the {\sc Gaia} location as an initial guess, is crucial in order to obtain high accuracy for the resultant extracted light curves. For these reasons, it is advantageous for the PSF centroid and broadness to be free parameters.

We note that star-specific deviations from the general trends seen for the centroid locations of all stars shown in Fig.~\ref{fig:centroid} could be caused by: (i) high proper motion stars; (ii) potential wobbling in the centre of light for an astrophysical reason (e.g. the changing light ratio between primary and secondary stars in an eclipsing binary system during the orbital phase; see \citealt{SouthworthBowman2025}); or (iii) rotational modulation, non-radial pulsations, and/or convective granulation changing the flux distribution on the surface of a star (see e.g. \citealt{Chiavassa_2011}). We do not rule out any of these effects except for granulation since this typically occurs in red (super)giant stars, which are not present in our sample. Hence we do not find any clear evidence for any astrophysical cause of the scatter in Fig.~\ref{fig:centroid} in our analysis. This is likely because such effects are much weaker than other instrumental sources of scatter in TESS image data. For example, we cross matched the measured x- and y-pixel centroid displacement patterns of all our stars with known high-proper motion stars and found no correlation. Moreover, stars with larger or atypical deviations in their measured centroids generally have larger centroid uncertainties and correspond to stars with fainter TESS and {\sc Gaia} $G$-band magnitudes, making it challenging to infer any astrophysical origin.

\subsubsection{Limitations}\label{sec: limitations}

The nature of a PSF-based approach, specifically that the flux of a target is extracted based on a 2D weighting function, means that not all of the flux of a source is necessarily utilised. This means that PSF-extracted light curves tend to include larger relative fractions of Poisson noise relative to SAP light curves. Primarily, this is due to smaller number of photons being used in PSF fitting of crowded sources compared to SAP light curves with large aperture masks. The smaller average fluxes in PSF light curves can be thus more susceptible to read-out noise and instrumental effects, including intra-pixel effects. In general, a SAP light curve with a larger aperture mask tends to average out systematic effects at the cost of increasing contamination (see \citealt{Papics2017a, BowmanMichielsen2021}). Whereas a PSF light curve reduces contaminated signal and follows the centroid of the PSF more accurately, at the cost of potentially introducing trends from the pixel-to-pixel sensitivity. Because this depends on the CCD properties of the region around the target, it is advisable to compare the light curves from both techniques for crowded regions.

A second major limitation of our PSF approach is that it may return unreliable results for (very) faint stars (i.e. $G \gtrsim 15$~mag), regardless of whether they are contaminated or not, since such stars are of similar brightness to the background in a TESS image. However, in practice this depends on a specific star's crowding and the time- and spatially dependent scattered light in TESS FFI data. Under these conditions, the fitting of a (very) faint star typically returns unrealistically high PSF broadness values, often extending beyond the image boundaries. Note that this limitation is set by the specifications of the TESS instrument rather than the PSF extraction technique itself, as we are pushing TESS mission data far beyond what it was originally designed to provide. Unfortunately, these criteria for faint stars yields an empirical brightness limit for successful targets of $G \lesssim 15$~mag in the SMC. This means that the majority of the BLOeM sample of SMC massive stars are not feasible for extracting reliable PSF light curves with TESS (see also Section \ref{sec: bloem phot}).

The quality of the final PSF light curve naturally also depends on the quality of the original TESS FFI data. In particular, it is well known that scattered light is a significant issue in TESS data \citep{Vanderspek+2018} and some software deals with it differently (see e.g. \citealt{RiddenHarper+2021} and \citealt{Roxburgh+2025}). We note that {\sc LEMONS} does not proactively discard bad quality data. However, it carries forward any flags placed by the TESS mission onto the final data product. In the case of severe scattered light, it is likely that the PSF fit does not converge. This triggers a warning and an additional {\sc LEMONS} flag to these data. Visual inspection of data is always advised as illustrated in the tutorial script\footnote{\url{https://github.com/pieterjanv314/Lemons/tree/main/tutorial}}. As such, the user has the freedom to decide whether any extracted light curves are meaningful for their science case.

\subsection{Example case studies}

After extensive testing and implementation, a sub-sample of 91 BLOeM targets have good quality TESS PSF light curves, which are indicated in Figs.~\ref{fig:HRD BLOeM} and \ref{fig:ra&dec BLOeM}. Although they comprise only about 10~per~cent of the full BLOeM sample, it is a larger sample than \citet{BowmanVanDaele2024} with our current study including 51 OB stars, 37 AF stars and 3 sgB[e] stars, thus a variety of stellar masses and evolutionary stages. For the stars where PSF photometry could not be performed, we also extracted SAP light curves using standard $3\times3$ aperture masks and provide these as a supplementary data product\footnote{\url{https://zenodo.org/records/20540863}}. However, we caution that these SAP light curves are heavily contaminated and should not be analysed blindly.


Here we demonstrate the advantage of our PSF approach compared to SAP methods using example stars within the BLOeM sample of massive stars in the SMC. In each example of specific scenarios, we compare the SAP and PSF light curves and demonstrate the effectiveness of the PSF methodology.

\subsubsection{BLOeM 4-058: False positive eclipsing binary}

\begin{figure}
    \centering
    \includegraphics[width=\columnwidth]{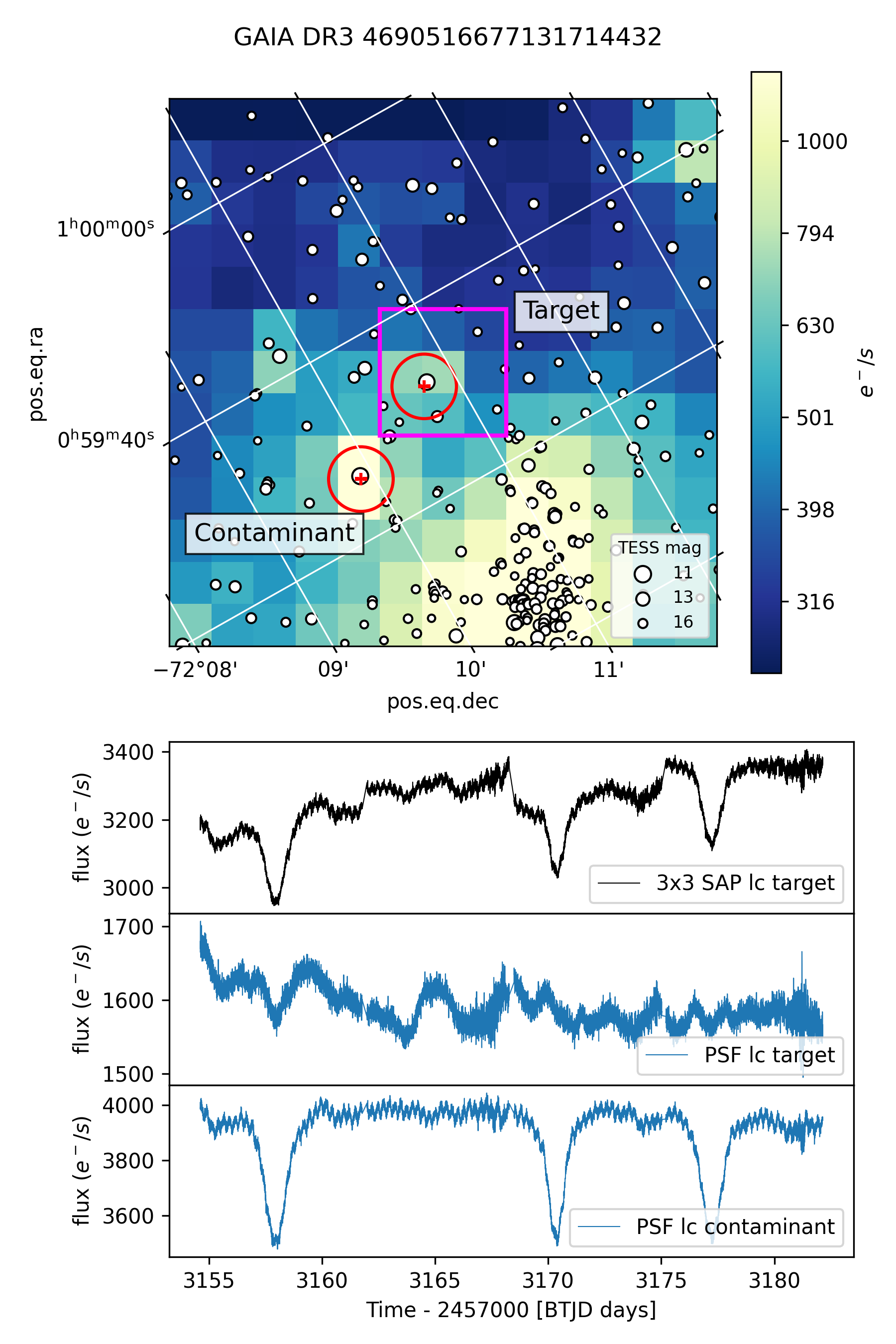}
    \caption{Demonstration of successfully disentangling contamination to reveal a false positive eclipsing binary with TESS data, sector 68. Upper panel: TESS pixel image around BLOeM 4-058 (GAIA DR3 4690516677131714432). The red circles indicate the FWHM of the Gaussian PSF fit and the magenta rectangle illustrates a standard $3\times3$ SAP mask. Lower panels: SAP (black) and PSF (blue) extracted light curves as well as the PSF light curve of contaminating stars. The SAP light curve is a combination of multiple sources of variability whereas the PSF light curves effectively disentangle them.}
    \label{fig:PSF lc}
\end{figure}

Figure~\ref{fig:PSF lc} shows the light curve of BLOeM 4-058 (Sk 80; GAIA DR3 4690516677131714432), which is a $\sim60~$M$_\odot$ O7Iaf$^+$ supergiant \citep{BLOeM2024, Bestenlehner+2025}. The SAP light curve of this star reveals SLF variability, higher frequency coherent pulsations as well as eclipses caused by a binary companion. However, the PSF light curve for this star, and its close neighbour, effectively disentangles all of these signals and successfully attributes the eclipses to the contaminating star and not the BLOeM target of interest (i.e. BLOeM 4-058). We identify the contaminating signal to come from HD\,5980 (GAIA DR3 4690516883290136832), a well known multiple system outside the BLOeM sample (see e.g. \citealt{Koenigsberger+2014,HD5980_2019,HD5980_2021}). It is worth noting that the contaminating source is about six times brighter than the BLOeM target of interest. This illustrates the importance of our PSF method because SAP light curves of massive stars can easily lead to false positive eclipsing binary systems (see \citealt{Abdul-Masih+2016, Prsa+2022}). This is especially important for massive stars in the LMC and SMC, because of the (moderate-to-severe) crowding, as can be seen in the TESS CCD image in the top panel of Fig.~\ref{fig:PSF lc}. 

\subsubsection{BLOeM 2-065: PSF light curve of an isolated constant star}

The jitter in addition to the systematic movement of the PSF centroids of all BLOeM stars (cf. Fig.~\ref{fig:centroid}) have an uncertain origin and could be caused by both astrophysical and/or instrumental effects, as discussed in Section \ref{sec:PSF for variable stars}. This could be a point of concern, since no artificial variability should be introduced into a light curve caused by centroid jitter. 

\begin{figure}
    \centering
    \includegraphics[width=\columnwidth]{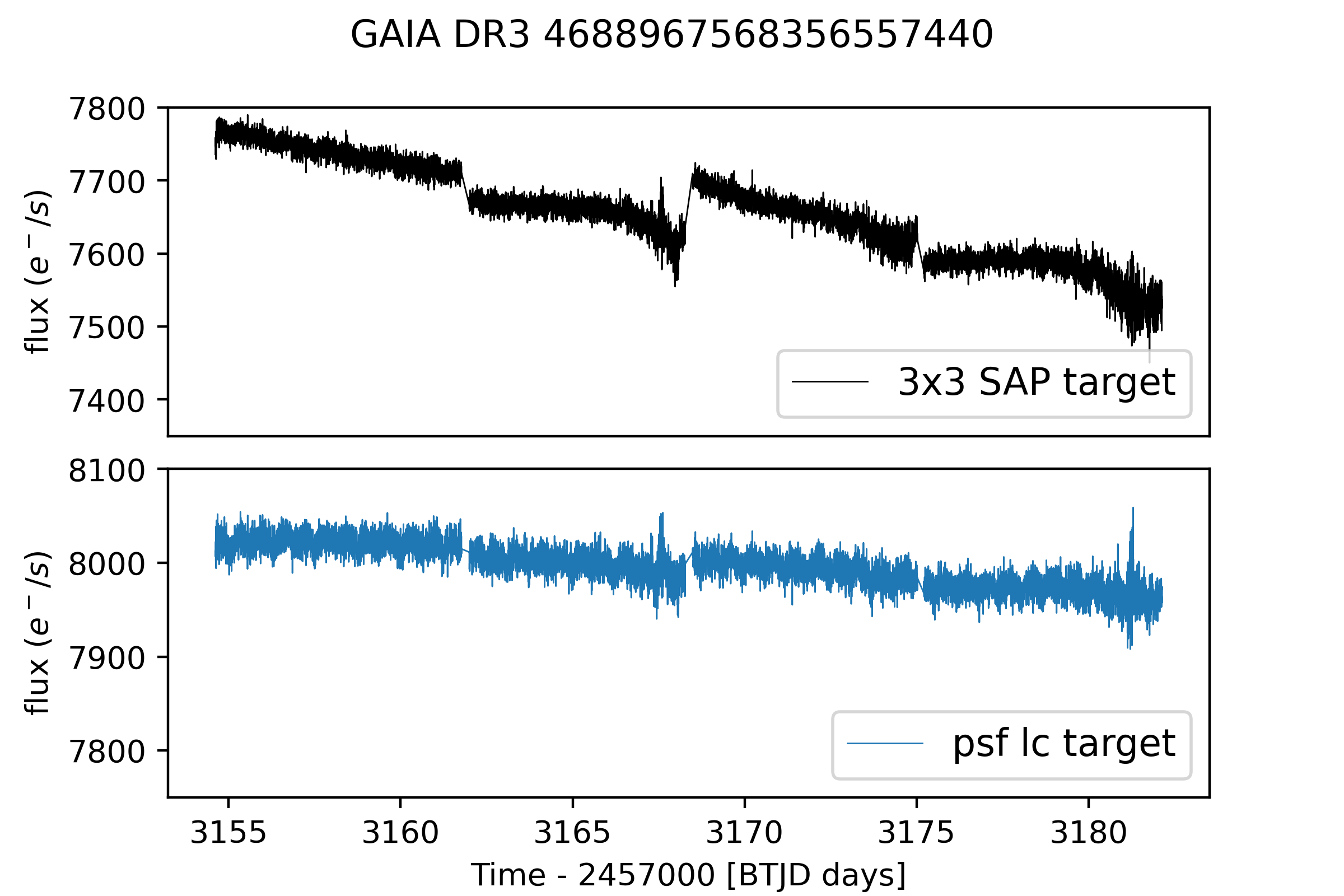}
    \caption{Demonstration of similar noise properties (i.e. jitter) between the $3\times3$ SAP light curve and the PSF light curve in TESS sector 68 for BLOeM 2-065 (AzV 121; GAIA DR3 4688967568356557440).}
    \label{fig:PSF lc constant star}
\end{figure}

To verify the impact of centroid jitter in a PSF light curve, we compare the SAP and PSF light curve for a F5 supergiant with constant brightness, BLOeM 2-065 (AzV 121; \citealt{BLOeM2024,Patrick+2025}), in Fig.~\ref{fig:PSF lc constant star}. This example demonstrates that having the PSF centroid as a free parameter in the fit does not introduce additional non-astrophysical variability in the light curve. Moreover, a decreasing trend and jumps are present in the SAP light curve which are far less severe in the PSF light curve. Since both light curves are not detrended and underwent the same background subtraction, this difference arises solely from the telescope drift and slight differences in pointing before and after the mid-sector data downlinking of the TESS spacecraft. In our PSF approach, this is mitigated by allowing for centroid freedom in the PSF fit. Whereas, in the SAP light curve extraction, a changing fraction of photons are being captured by the aperture mask, which is fixed in terms of pixel location. This example therefore also illustrates the advantage of a optimised flexible aperture mask in the PSF fitting approach.

\subsection{TESS and OGLE comparison}

OGLE is a long-term ground-based observational campaign program that assembles time-series photometry of Galactic, LMC, and SMC stars \citep{OGLE_IV_2015,OGLE_SMC_2023,Glowacki+2024}. Although OGLE is unable to match TESS's photometric precision, nor its high duty cycle and short cadence, its smaller pixel size and observing strategy allows for a much higher spatial resolution and long base lines of several years. This means that OGLE photometry is entirely complementary to TESS light curves. 

\begin{figure}
    \centering
    \includegraphics[width=\columnwidth]{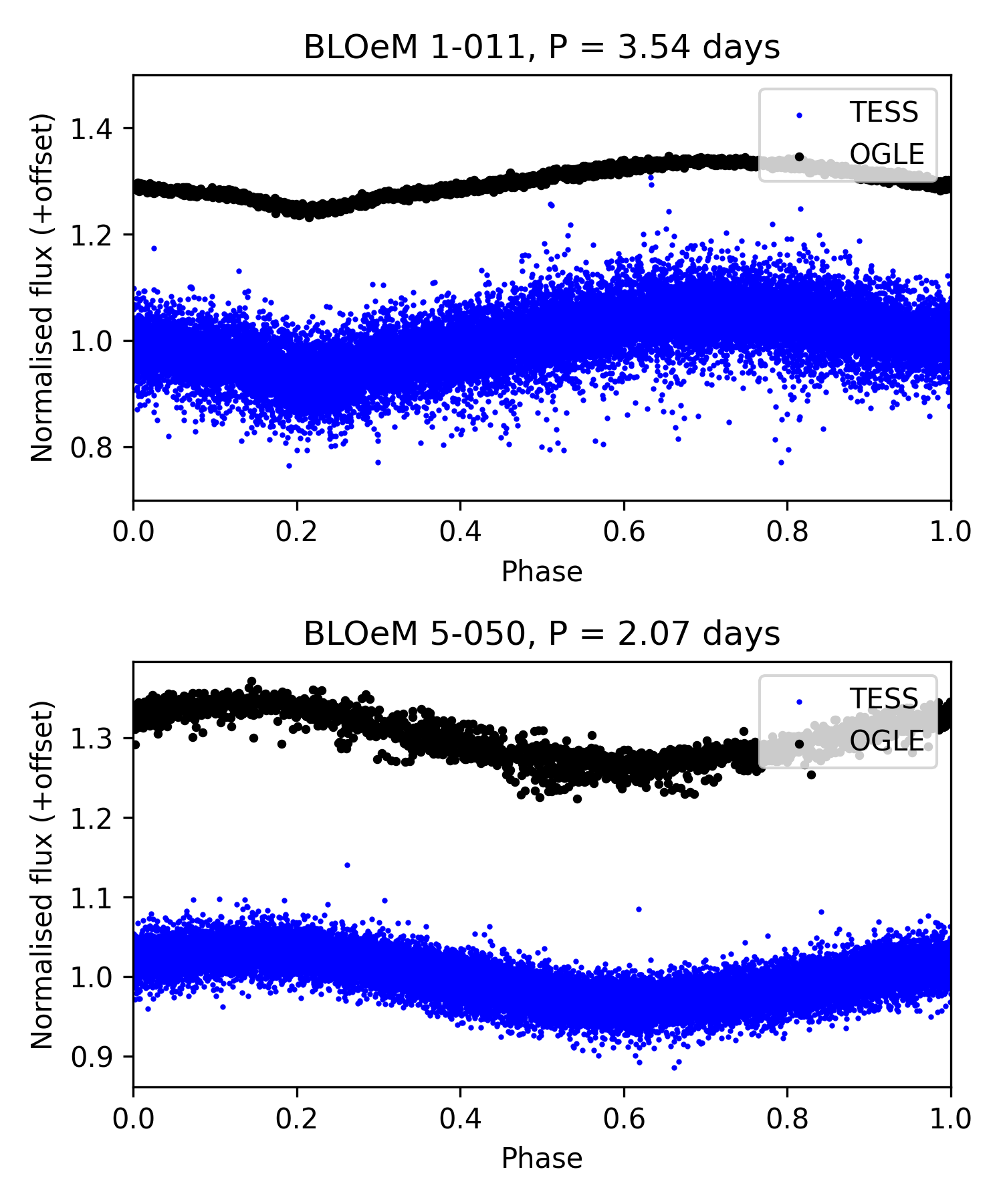}
    \caption{Comparison of TESS PSF and OGLE $I$-band light curves for two known binaries. The OGLE data is taken from \citet{Glowacki+2025_OGLEcat}. The top and bottom panels show the phase folded OGLE (black, with offset) and PSF TESS (blue) light curves for BLOeM 1-011 (GAIA DR3 4690506712804881792) and BLOeM 5-050 (GAIA DR3 4687501953697501568), respectively. The orbital period, indicated above each plot, extracted from TESS and OGLE are comparable and is recovered robustly.}
    \label{fig:OGLE}
\end{figure}

To verify that {\sc Lemons} produces legitimate light curves, we compared our extracted TESS light curves to OGLE light curves obtained from \citet{Glowacki+2025_OGLEcat} of known variable stars. Figure~\ref{fig:OGLE} shows a comparison of TESS and OGLE light curves for two binary systems in our sample, BLOeM 1-011 and 5-050. Both light curves are phase folded on the OGLE extracted binary period. Despite a sometimes larger scatter with TESS data, the period recovery is excellent. For BLOeM 1-011, we find a TESS period of $3.5270 \pm0.0004$~d which is comparable with the period $3.5358997 \pm 0.0000003$~d extracted from OGLE photometry. The situation is similar for BLOeM 5-050, with a TESS period of $2.0780\pm0.0001$~d and an OGLE period of $2.0731282 \pm 0.0000002$~d. Formal errors for both TESS and OGLE are from estimated based on a least-squares fit to each respective light curve, but are likely underestimates of the true uncertainties. This overall demonstrates that our PSF method indeed captures the astrophysical signal from the target of interest within the crowded TESS data.

\section{Photometric variability of SMC massive stars in the BLOeM sample}
\label{sec: bloem phot}

PSF light curves were successfully extracted for 91 BLOeM stars, with a total of 281 light curves of indivdual sectors (at the time of analysis, FFIs for sectors 27, 28, 67 and 68 were available that included the SMC). Although this is only about 10~per~cent of the total BLOeM sample, each individual light curve is an accomplishment considering that this technique pushes the possibilities of what was thought achievable with TESS mission data --- the instrument was not designed to study SMC stars. In this faint and crowded regime, the success of PSF fitting depends largely on the brightness ratio and local image geometry (i.e.\ location of stars on the 2D sky projection on the CCD) of the target with respect to nearby stars in the CCD image. The PSF approach typically starts to fail for stars with $13 \lesssim G \lesssim 14~\rm{mag}$ and/or the ratio of the target's flux relative to the total flux of all stars (from the TESS Input Catalogue) within a 42~arcsec radius is less then than 40~per~cent (i.e. $F_{*}/F_{42^{''}} \leq 0.4$), see also Section \ref{sec: limitations}. However, in cases with little contamination, we have successfully reached the brightness limit of $G\sim15$~mag. Beyond this point, targets are indistinguishable from background noise in the TESS CCD images.

We note that not every available sector of a target yielded a reliable (i.e. astrophysically dominated) PSF light curve. This is often due to data issues marked by the TESS quality flags within a specific sector (e.g. scattered light) as mentioned in section~\ref{sec: limitations}. But, even with acceptable quality flags, the success of PSF fitting might differ from one sector to the next. This is expected given the fine margins of TESS data we are exploiting. For example, additional background noise or rotated positions of the stars on the image, which results in two stars sharing a pixel where previously they did not, could be enough for a specific sector to fail in extracting a PSF light curve. 

Because of the technical limitations of the TESS instrument, there is a clear observational bias of being able to successfully extract a PSF light curve for intrinsically more luminous stars. In the HR~diagram in Fig.~\ref{fig:HRD BLOeM}, the successfully extracted 91 stars are overplotted as blue diamonds. We also include histograms adjacent to the x- and y-axes that show the completeness of stars with a PSF light curves compared to the complete BLOeM sample as a function of $T_{\rm eff}$ and $L$, respectively. Therefore, our sample with successful PSF light curves is dominated by main sequence O dwarfs (i.e. masses above $\sim 20$~M$_{\odot}$) and post-main sequence stars (i.e. giants and supergiants). 

Regardless of the relatively low success rate, our PSF light curves reveal several types of variability indicative of coherent pulsation modes excited by a heat-engine mechanism, eclipsing binaries, SLF variability, and rotational modulation caused by chemical abundance spots (i.e. candidate magnetic stars), which are discussed in greater detail in the following subsections. We also provide a variability classification for each star with a successfully extracted PSF light curve in Table~\ref{tab:psf-lc}.

\subsection{Pulsations}
The lower metallicity of the SMC compared to the Galaxy means there is less opacity in the iron-bump at 200\,000~K, which is the excitation mechanism for coherent pulsation modes in massive stars \citep{DziembowskiPamyatnykh1993}. Therefore, it is expected that fewer, if any, coherent heat-driven pulsating massive stars exist in the SMC (e.g. \citealt{Salmon+2012}). However, Be stars, for which pulsations can be triggered by stochastic convection and are related to their fast rotation rates (e.g. \citealt{Neiner+2020, Labadie-Bartz2022}), seem to be more likely to exhibit pulsations than ordinary B‑type stars in low-metallicity environments (see \citealt{Diago+2008}).

On the other hand, IGWs are predicted by numerical simulations to be stochastically excited at the interface of convective and radiative zones \citep{Rogers2013, Rogers+2017, Edelmann+2019ApJ, Thompson+2024_3Dsim}, and have been inferred to be one of the causes of SLF variability in massive stars \citep{Bowman+2019_nat, Bowman+2020aa}. IGWs are especially important to study in low-metallicity massive stars, since such stars may lack entirely sub-surface convection zones \citep{Jermyn+2022}. This means that either: (i) no SLF variability should be observed for such stars and thus it must solely be caused by subsurface convections (see e.g. \citealt{Cantiello+2009,Cantiello+2021}), or (ii) SLF variability must, at least partly, arise from IGWs excited by the convective core (see \citealt{Bowman2023}). For example, \citet{BowmanVanDaele2024} found little difference in the morphologies of SLF variability across Galactic, LMC and SMC massive stars. Their sample spanned a range of masses and ages, and specifically included stars both inside and outside of the transparency windows for sub-surface convection zones \citep{Jermyn+2022}, with little difference in the amplitudes or frequencies of SLF variability. 

Below we discuss two cases of pulsational variability: one star with coherent pulsations and one with SLF variability. We find a total of 14 coherent pulsator (candidates) and
numerous examples of SLF variability in our sample of massive stars with PSF light curves, with the latter being discussed in more detail in Section~\ref{sec:SLF}. 

\subsubsection{BLOeM 1-040: coherent heat-driven pulsations in an Oe/Be star}

\begin{figure}
    \centering
    \includegraphics[width=\linewidth]{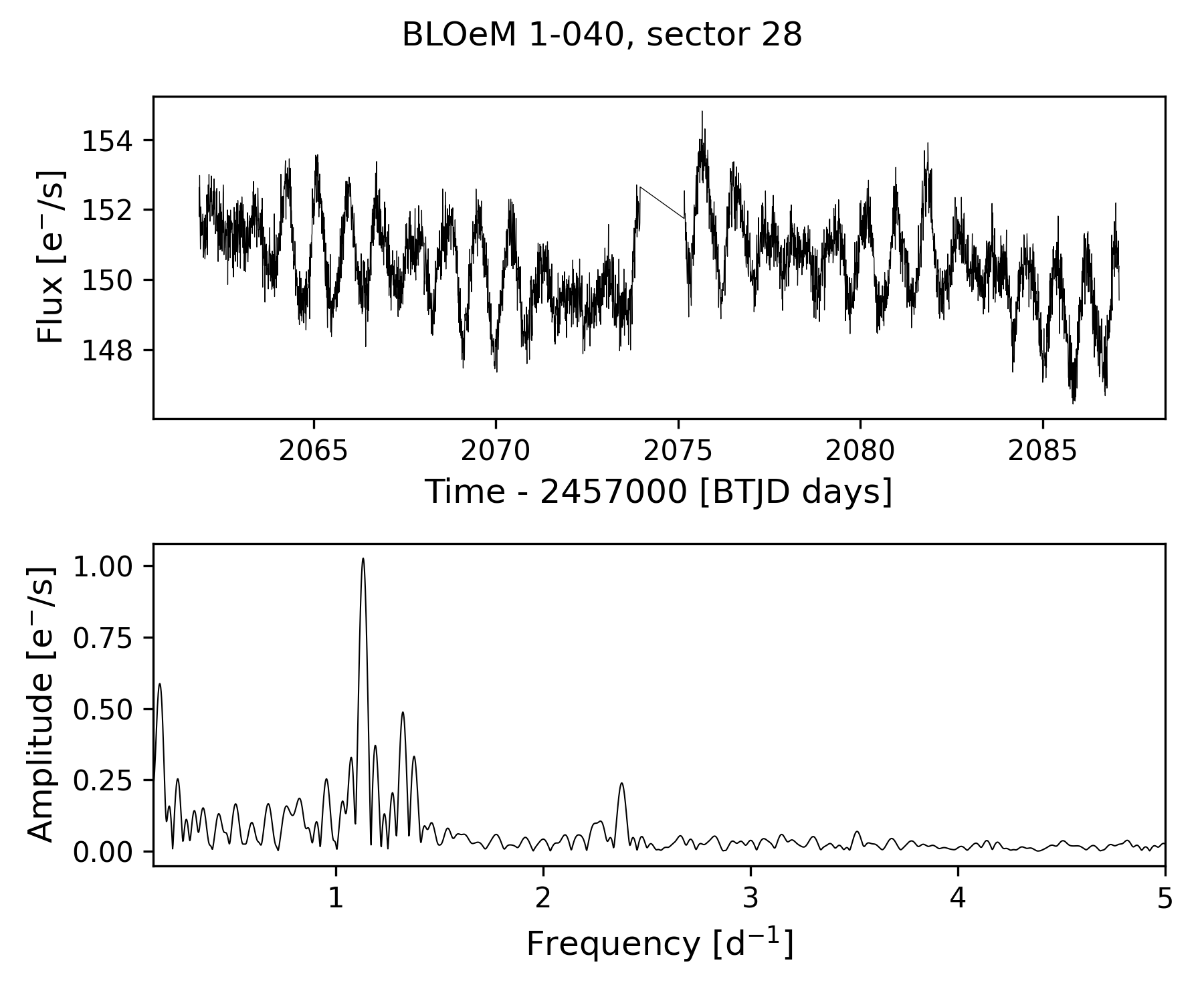}
    \caption{PSF light curve and Lomb-Scargle periodogram of BLOeM 1-040 (GAIA DR3 4690525438865707904), which has gravity-mode pulsations typical of an Oe/Be star.}
    \label{fig: GAIA DR3 4690525438865707904}
\end{figure}

BLOeM 1-040 (GAIA DR3 4690525438865707904) is an O9.7\,III:ne star and a known SB1 system with an orbital period of approximately 45~d \citep{Bodensteiner+2025}, which is longer than a single TESS sector. Our PSF light curve and the corresponding Lomb-Scargle periodogram for sector~28 are shown in Fig.~\ref{fig: GAIA DR3 4690525438865707904}. 

We detect a dominant variability period in our PSF light curve with a frequency of 1.13~d$^{-1}$. Since this star is an Oe/Be star, and such stars are commonly pulsators (see \citealt{Labadie-Bartz2022}), we conclude that this star has gravity-mode pulsations (see also \citealt{StankovHandler2005, Burssens+2020}). There is also a harmonic of the dominant frequency at 2.26~d$^{-1}$, which further supports this conclusion as frequency groups such as this are common in pulsating Be stars (see \citealt{Kurtz2015b, Labadie-Bartz2022}). \\
\\

\subsubsection{BLOeM 1-009: SLF variability in a supergiant}

\begin{figure}
    \centering
    \includegraphics[width=\linewidth]{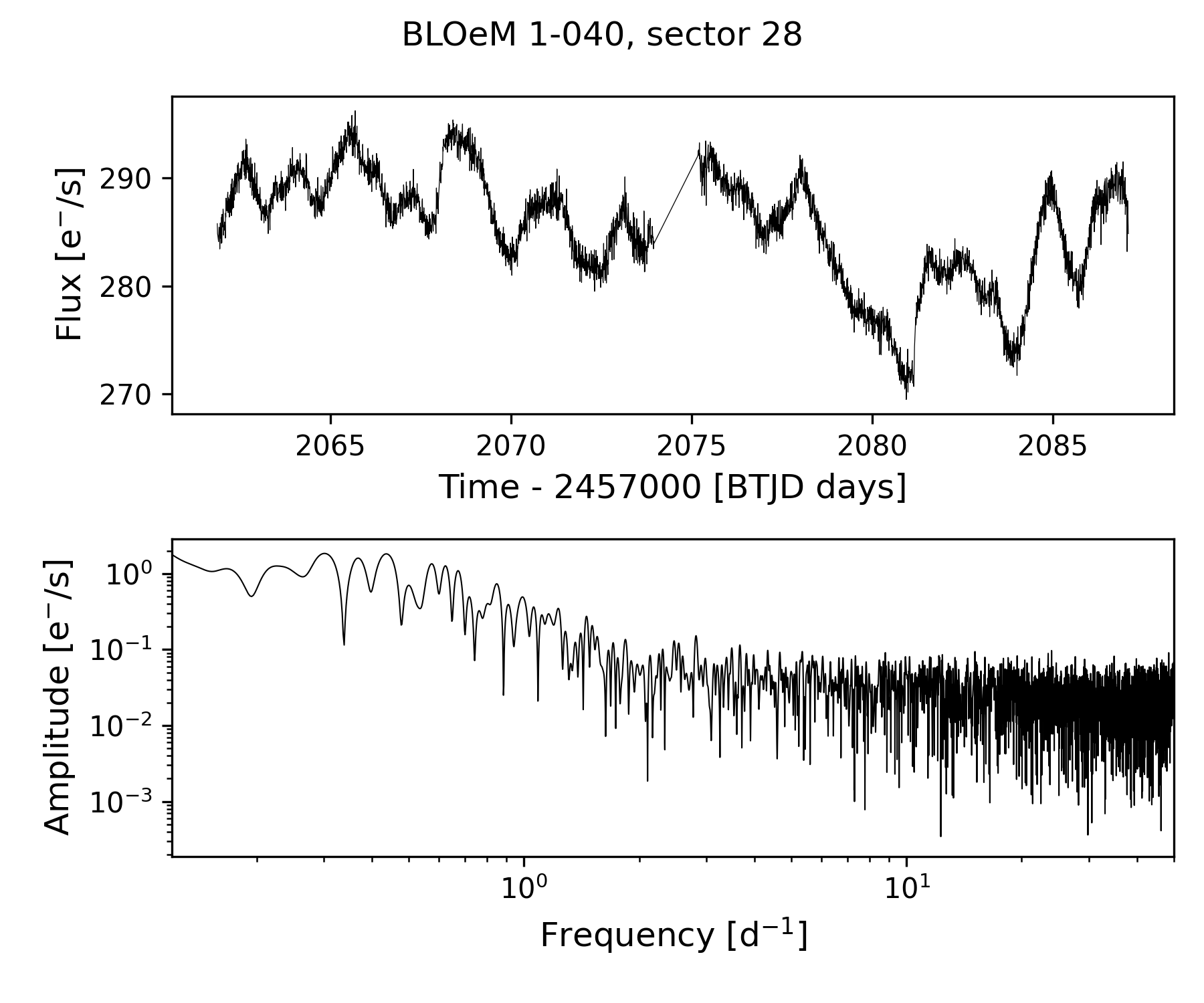}
    \caption{PSF light curve and Lomb-Scargle periodogram of BLOeM 1-009 (GAIA DR3 4690520387983090432), which is a typical example of a massive star with SLF variability.}
    \label{fig: GAIA DR3 4690520387983090432}
\end{figure}

BLOeM 1-009 (GAIA DR3 4690520387983090432) is a B1\,Ia supergiant \citep{BLOeM2024,Britavskiy+2025} and its light curve is a typical example of massive star SLF variability. Our PSF light curve and the corresponding Lomb-Scargle periodogram for sector~28 are shown in Fig.~\ref{fig: GAIA DR3 4690520387983090432}, which clearly illustrate the quasi- and multi-periodic nature of SLF variability. 

Since SLF variability is common in supergiants \citep{Bowman+2019_nat}, it creates additional difficulties for analysing spectroscopic data of such stars. This is because pulsations change the shapes of spectral lines and introduce significant radial velocity (RV) scatter on time scales of hours-to-days (see \citealt{Aerts2009b, AertsRogers2015, Kalita+2025}). Therefore, generally speaking, pulsating stars need about twice as many spectroscopic epochs to disentangle their binary periods from potentially similar pulsation periods (see \citealt{SouthworthBowman2025}). This star (and others like it) demonstrate the complementarity of using time-series photometry to quantify the contribution of RV variability caused by pulsations when interpreting multi-epoch spectroscopy to identify binaries.

\subsection{Indications of binarity and rotation}
\label{sec: binaries}

\begin{table*}
\centering
\caption{Summary of stars with a PSF light curve extracted in this work that indicate potential binarity. Categories for the TESS variability are ellipsoidal variability (EV), rotational modulation (RotMod), stochastic low frequency variability (SLF) and eclipsing binary (EB). Uncertainties for the dominant photometric period are the formal $1\sigma$ values calculated using a least-squares fit to the light curve. BLOeM binary properties are from {\color{blue} Ovadia et al.} (in prep.) for the O type stars and \citet{Britavskiy+2025}, {\color{blue}Katabi et al.}  (in prep.) for the B type stars.}
\begin{tabular}{c c c c c c}
\hline
BLOeM ID & SpT & TESS variability & BLOeM binary status & BLOeM binary period & PSF light curve period \\
& & & & (d) & (d) \\
\hline
1-011 & B1.5\,II & EV or RotMod & SB2 & 7.07 & 3.5270 $\pm$ 0.0004 \\
1-106 & B1.5\,e & SLF (eclipses?) & $-$ & $-$ 
& 1.00485 $\pm$ 0.00006 \\
3-042 & O6\,I(f) & EV or RotMod & SB2 &  2.32  & 1.15909 $\pm$ 0.00005\\
5-050 & O9.7\,V: & EV or RotMod & SB2 &  4.17  & 2.0780 $\pm$ 0.0001\\
6-105 & O6\,V:n & EV or RotMod & SB1 & 13.8  & 0.70356 $\pm$ 0.00009\\

8-006 & B2.5\,II-Ib & EV or RotMod (or SLF?) & SB1/LPV & 0.82/4.7 &  4.823 $\pm$ 0.002\\
8-097 & F0: & contaminated (EB?) & $-$ & $-$
& 1.5642 $\pm$ 0.0004 \\
\hline
\end{tabular}
\label{table:binaries}
\end{table*}

Among the 91 BLOeM targets with successful PSF extracted TESS light curves, there are seven targets that show eclipses or ellipsoidal variability (EV), which are photometric signatures of binarity. It is often difficult to differentiate EV from rotational modulation, and impossible in a synchronised binary system without any extra information (see \citealt{SouthworthBowman2025} for a review). However, finding a photometric variability period in agreement with a binary period found with BLOeM spectroscopy is therefore a strong confirmation of binarity.

For two of the seven candidate binaries in our work, BLOeM 6-105 and 8-006, the dominant photometric period does not correspond with the binary period based on BLOeM spectroscopy. Assuming the BLOeM binary period is correct, this discrepancy could be caused by: (i) the variability detected in TESS photometry arises from rotational modulation rather than binarity in a non-synchronous system; (ii) it is also possible that the photometric signatures of the BLOeM binary period is too low in terms of S/N to be detected in our TESS light curves; and (iii) photometric contamination from nearby sources in the TESS CCD images leading to a false positive.

Unfortunately, an unambiguous spectroscopic binary period has not yet been determined for two of these seven stars, BLOeM 1-106 and 8-097. Therefore, in this work we provide photometric periods to assist in solving these candidate binary systems in the future.

All seven photometric binary candidates found in this work are presented in Table~\ref{table:binaries}, and phase folded TESS PSF light curves for each star are presented in Fig.~\ref{fig: bin_spec} and \ref{fig: bin_nospec}. Periodograms of each of these seven stars is provided in Appendix~\ref{fig: appendix: binaries}. Below, we provide some comments on the individual targets.

\begin{figure}
    \centering
    \includegraphics[width=\linewidth]{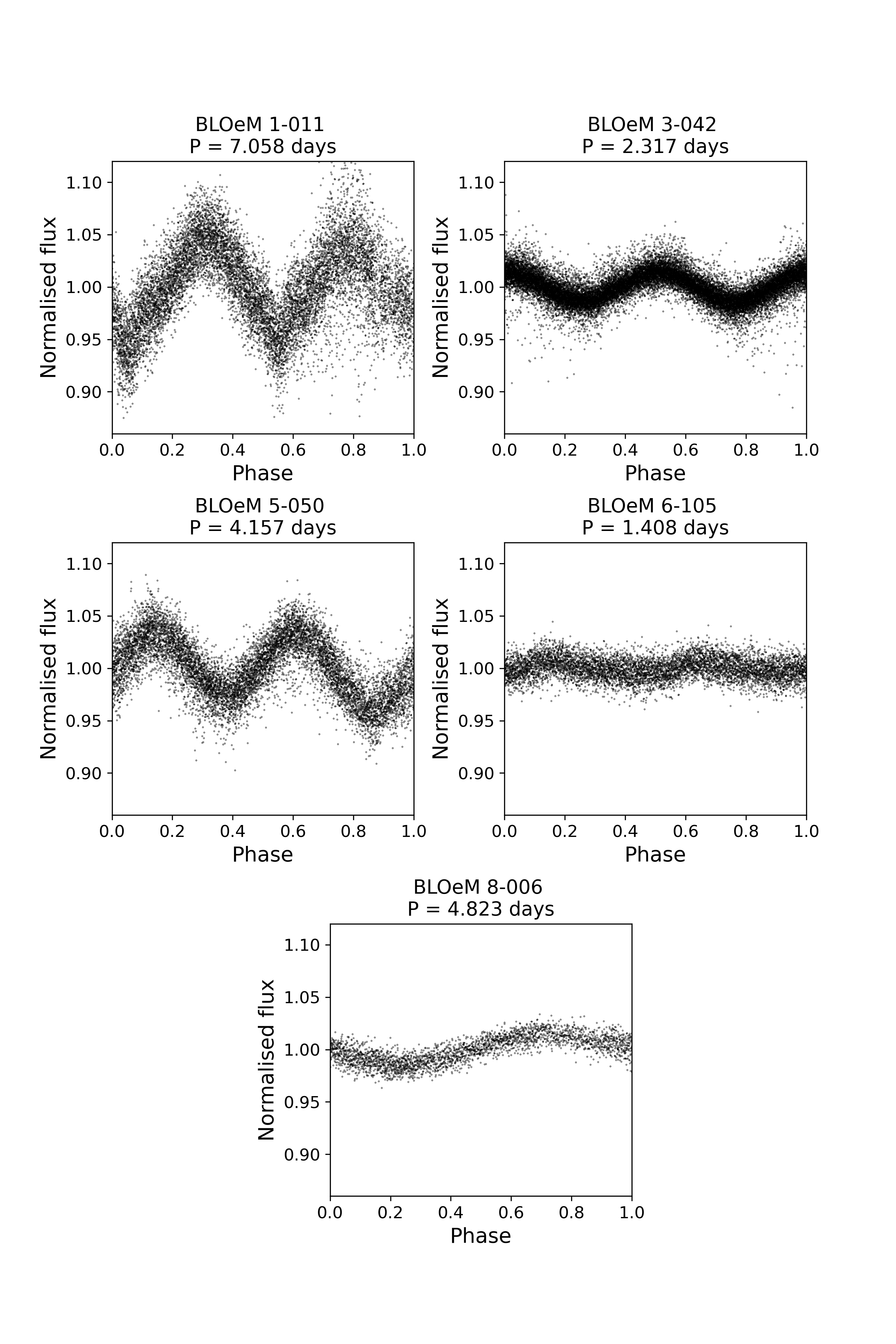}
    \caption{
    Phase folded light curves of targets with spectroscopically confirmed binary periods which agree with our PSF TESS light curves. The BLOeM identifier and binary period, $P$, are mentioned in each subplot's title. The vertical axis has the same scale for all plots for comparison reasons. }
    \label{fig: bin_spec}
\end{figure}

\begin{figure}
    \centering
    \includegraphics[width=\linewidth]{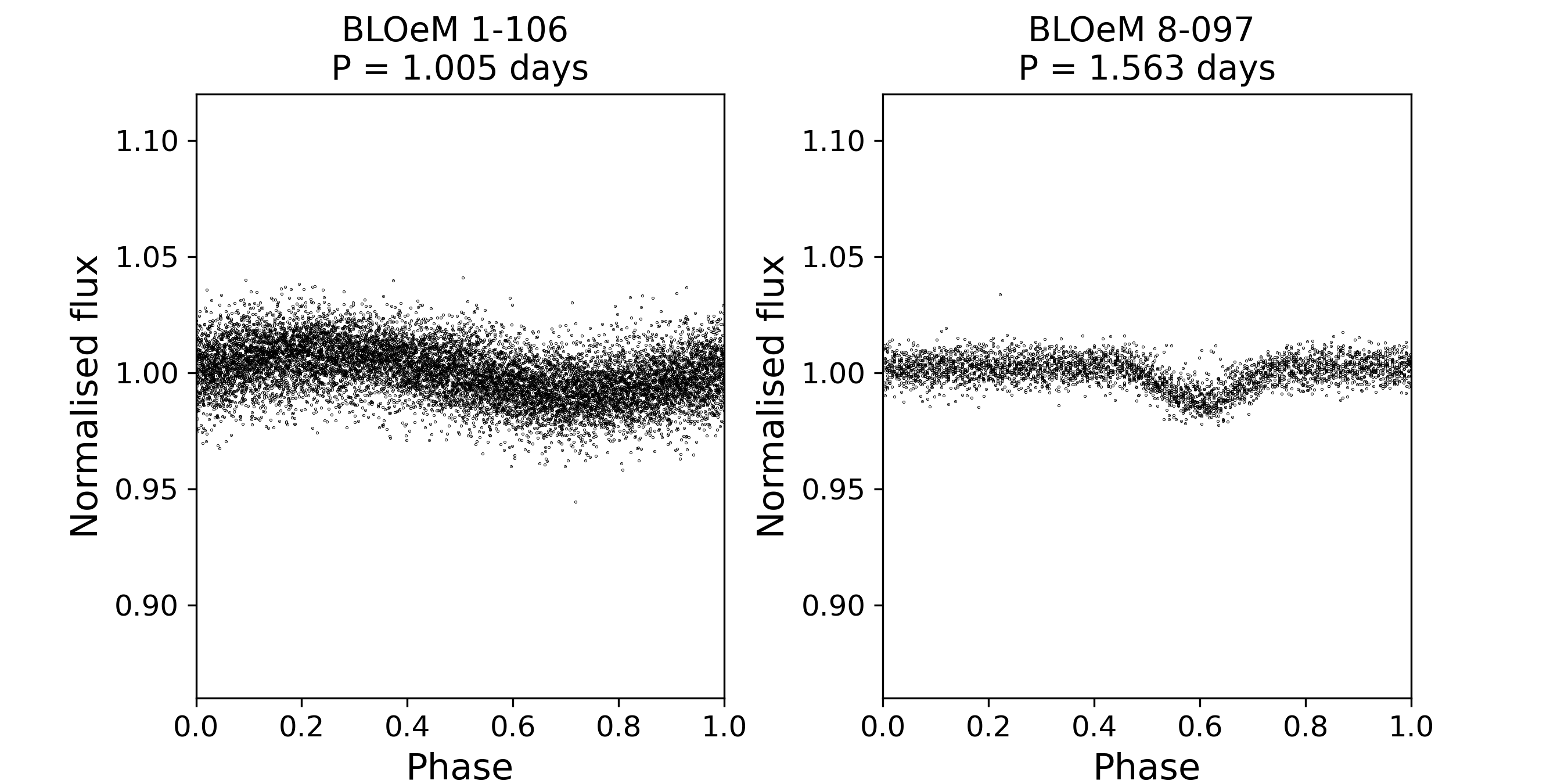}
    \caption{
    Phase folded light curves of targets with photometric indications of binarity or rotation, but without spectroscopic confirmation. The BLOeM identifier and best-fitting photometric period, $P$, are mentioned in each subplot's title. The vertical axis has the same scale as in Fig.~\ref{fig: bin_spec}} for comparison reasons.
    \label{fig: bin_nospec}
\end{figure}

\subsubsection{BLOeM 1-011}
BLOeM 1-011 (GAIA DR3 4690506712804881792) is a known SB2 system, with a primary star of spectral type B1.5\,II \citep{BLOeM2024,Bestenlehner+2025} and binary period of $7.07$~d \citep{Britavskiy+2025, Sana+2025NatAs}. The dominant photometric period in the TESS light curve is $3.54$~d. However, this is half of the spectroscopic binary period found with BLOeM spectroscopy \citep{Britavskiy+2025}, which is a strong confirmation that we are observing the same signal. Therefore, we opted to phase fold the PSF TESS light curve using the BLOeM period, which is shown in Fig.~\ref{fig: bin_spec}. Assuming that $7.07$~d is the true binary orbital period, the symmetry of the two dips in the phase folded light curve suggests that this is a binary system with EV and with stars of similar brightness. 

\subsubsection{BLOeM 1-106}
The rapidly rotating B1.5\,e star BLOeM 1-106 (GAIA DR3 4687505801988031232; \citealt{BLOeM2024,Bestenlehner+2025,Bodensteiner+2025}) has photometric variability with a dominant period of $1.01$~d in our PSF TESS light curve (see Fig.~\ref{fig: bin_nospec}). To date, there is no spectroscopic binary period established for this star from BLOeM spectroscopy. There is also a weaker 5-d period in the light curve, however we deem it to be likely arising from contamination since it is not present in the central pixel where one would expect the signal to be strongest if arising from the target of interest. Such a short period of 1.01~d is unlikely to be a binary period, so we infer that this period is more likely to be rotational modulation, which is sufficiently fast to be consistent with its spectroscopic classification as a Be star.

\subsubsection{BLOeM 3-042}
BLOeM 3-042 (Sk~18; GAIA DR3 4685947553531340800) is a known O6\,I(f)+O7.5 SB2 binary system \citep{BLOeM2024, Sana+2025NatAs}. TESS photometry reveals variability with a dominant period of $1.16$~d. {\color{blue}Ovadia et al.} (in prep.) determined a spectroscopic binary period of 2.32~d, which is twice the dominant photometric period. Therefore, we conclude that the photometric variability is caused by binarity, rather then rotation. We opted to phase fold the TESS light curve using double the dominant TESS period in Fig.~\ref{fig: bin_spec} to be in line with spectroscopic binary period.

\subsubsection{BLOeM 5-050}
BLOeM 5-050 (GAIA DR3 4687501953697501568) is a known O9.7\,V: + early-B SB2 binary system \citep{BLOeM2024,Bestenlehner+2025,Sana+2025NatAs}. TESS photometry reveals variability with a dominant period of $2.08$~d. {\color{blue}Ovadia et al.} (in prep.) found a spectroscopic binary period of 4.17~d, which is twice the dominant photometric period. This indicates that the photometric variability is EV, rather then rotational modulation. We opted to phase fold the TESS light curve by double the dominant TESS period in Fig.~\ref{fig: bin_spec} to be in line with spectroscopic binary period.

\subsubsection{BLOeM 6-105}
BLOeM 6-105 (GAIA DR3 4686410997616777984) is a known SB1 system with spectral type O6\,V:n  \citep{BLOeM2024, Bestenlehner+2025, Sana+2025NatAs}. {\color{blue}Ovadia et al.} (in prep.) determined a spectroscopic binary period of 13.8~d. Such a long spectroscopic period is challenging to detect in the light curve of a 27-d TESS sector. However, our PSF light curve (see Fig.~\ref{fig: bin_nospec}) shows weak variability at a period of $0.7$~d. This is very short for a massive binary period and therefore we surmise it is caused by rotational modulation. 

\subsubsection{BLOeM 8-006}
BLOeM 8-006 (GAIA DR3 4689037181200454656) is a B2.5\,II-Ib star \citep{BLOeM2024,Bestenlehner+2025} with a reported spectroscopic periodicity at 0.82~d by \citet{Britavskiy+2025} based on the original nine BLOeM epochs. However, {\color{blue}Katabi et al.} (in prep.) find a revised orbital period of 4.65~d using all 25 BLOeM spectroscopic epochs. We note that both studies retrieved a different significant period based on a Lomb-Scargle periodogram analysis, but with additional epochs the longer orbital period is preferred. 

Our PSF TESS light curve of sector 27 does not reveal any significant signal at $P=0.82~$d. Instead, we find a dominant period of 4.82~d (see also Fig.~\ref{fig: bin_spec}). Given that our photometric period is about one sixth of the duration of a TESS sector, the true uncertainty on this period is likely much larger than the formal statistical error based on a least-squares fit to the light curve reported in Table~\ref{table:binaries}. Additionally, we also find that there is sector-to-sector scatter for such large periods. For example, repeating our period determination analysis for sector 67 instead of sector 27 yields $P\simeq4.91$, thus demonstrating similar consistency with the longer BLOeM spectroscopic orbital period. The much worse accuracy compared to the formal precision is reflected by the broad peak in the periodogram of the TESS light curve (see Fig.~\ref{fig: appendix: binaries}). Nevertheless, the reported spectroscopic period by {\color{blue}Katabi et al.} (in prep.) is consistent with the value obtained from the periodogram of our TESS light curve. On the other hand, a period of 0.82~d is very small for a massive star in a binary system. This could represent a harmonic of EV detected in our light curve but a sixth harmonic is unexpected. Therefore, we conclude that the TESS photometric period of $4.8~$days, in agreement with the reported spectroscopic period by {\color{blue}Katabi et al.} (in prep.), is the orbital period and that the 0.82~d period reported by \citet{Britavskiy+2025} is an alias.

\subsubsection{BLOeM 8-097}
BLOeM 8-097 (GAIA DR3 4689058999614762368) is an F0 supergiant with unknown binary status \citep{Patrick+2025}. Its light curve, shown in Fig.~\ref{fig: bin_nospec}, reveals weak, but regular eclipses with a period of 1.56~d. These shallow (and somewhat asymmetric) eclipses imply binarity. However, given such a short period and such an evolved star, this seems unlikely. On the other hand, this star does have one of the lowest standard deviations for extracted RVs within the subsample of evolved BLOeM stars studied by \citet{Patrick+2025}. With such a faint signal, extra caution is advised due to the risk of contamination. Indeed, when investigating the light curves of the individual TESS pixels, the signal appears to be stronger in a pixel next to the target, where four faint star are located ($16.5 < G < 18.0$~mag), compared to the target ($G = 13.8$~mag). Unfortunately, all these stars are too faint to be captured within the PSF. However if one of them has relatively deep eclipses, this could still enter our PSF light curve of BLOeM 8-097. Hence, we deem the eclipses to be potential contamination.

\section{Stochastic low-frequency (SLF) variability at SMC metallicity}
\label{sec:SLF}

In addition to the study of coherent pulsators and binary systems, our new PSF TESS light curves allow us to investigate the nature of SLF variability in SMC stars and how its morphology correlates with stellar parameters, in particular the location in the HR~diagram. Moreover, since almost all studies of SLF variability to date have been for Galactic stars, in this case we can study massive stars in the SMC, and determine the impact of metallicity on SLF variability. We explore this topic by mapping the observed SLF variability for our subsample of stars with extracted PSF light curves, compare the results with what has been found previously for Galactic, LMC and SMC massive stars, and investigate any difference in SLF morphology among main-sequence and post-main sequence stars.

The most common approach to study SLF variability is to fit the periodogram (see review by \citealt{Bowman2023}), but other methods include fitting the light curve directly using Gaussian process regression (see \citealt{BowmanDW2022}). Inspired by the original large sample studies of \citet{Bowman+2019_nat} and \citet{Bowman+2019_corot}, we use a semi-Lorentzian function $\alpha(\nu)$, with $\nu$ being frequency, to fit the periodogram with least squares and characterise the maximum amplitude $\alpha_0$, characteristic frequency, $\nu_{\rm char}$, and steepness, $\gamma$, on top of a white noise level $C_{\rm w}$ using:

\begin{equation}\label{eq:SLF}
    \alpha(\nu) = \frac{\alpha_0}{1+\left(\frac{\nu}{\nu_{\rm char}}\right)^{\gamma}} + C_{\rm w} ~ .
\end{equation}

\begin{figure*}
    \centering
    \includegraphics[width=\linewidth]{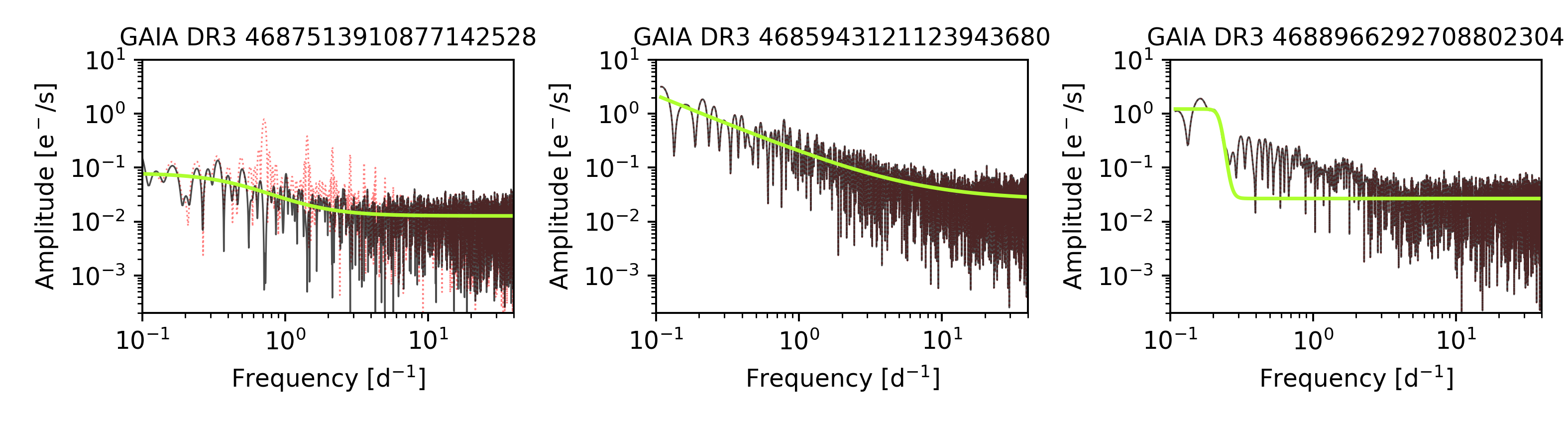}
    \caption{Example fitted SLF variability profiles for different stars. Left panel: example of a star that underwent prewhitening, with the original and residual periodograms shown in red and black, respectively. Middle panel: Fitted SLF variability profile that has been excluded form further analysis because $\nu_{\rm char} < 0.15$~d$^{-1}$. Right panel: Example of SLF variability with a poorly fitted profile and hence excluded from further analysis because of the visual discrepancy between the fitted curve and periodogram.}
    \label{fig: failed_SLF}
\end{figure*}

Light curves with significant periodicity (pulsations or eclipses) underwent iteratively prewhitened with {\sc Period04} \citep{period04_2005} to remove their dominant peaks in the periodogram as this would prevent a successful fitting. An example of this is shown in the left-hand panel of Fig.~\ref{fig: failed_SLF}. The resultant profiles of SLF variability are for periodograms of light curves spanning a single sector, since the large data gap in between TESS sectors would introduce an artificial excess at low frequencies in the periodograms. When multiple consecutive sectors are available of the same star, their light curves are treated as independent observations, but all underwent the same process separately.

An typical example of a star with SLF variability is shown in Fig.~\ref{fig: GAIA DR3 4690520387983090432}.
However, sometimes instrumental effects from TESS (see sections~\ref{sec: limitations} and \ref{sec: bloem phot}) can cause an additional excess at very low frequencies (i.e. $\nu < 0.1$~d$^{-1}$), resulting in a periodogram that keeps increasing towards lower frequencies and does not plateau at a discernible value of $\alpha_{0}$. An example of this behaviour is shown in the middle panel of Fig.~\ref{fig: failed_SLF}. In the cases when instrumental effects dominate astrophysical variability, it is difficult to reliably measure the properties of the SLF variability. Therefore, we filtered out stars with fitted SLF variability profiles with $\nu_{\rm char} < 0.15~\rm{d^{-1}}$. This is motivated by the fact that such small frequencies cannot be resolved in a single sector of TESS data, with the frequency resolution being proportional to the inverse of the light curve's time span. Based on this, we filtered out 139 sectors of light curves for 72~stars. Finally, after a visual inspection of the quality of the fits, we removed a further 17 sectors of light curves because the fitted SLF variability model clearly did not match the data. An example of a poorly fit star is shown in the right-hand panel of Fig.~\ref{fig: failed_SLF}. 

After this filtering process, we are left with a total of 105 sectors for 51 massive stars, for which there is a high degree of confidence that the SLF variability is astrophysical. For all of these stars, the Bayesian Information Criterion (BIC; \citealt{BIC}) was used to check that best fitted Lorentzian profile (cf. Eqn~\ref{eq:SLF}) is statistically preferred over the simpler model of a horizontal line (i.e. zero gradient white noise only).

\begin{figure*}
    \centering
    \includegraphics[width=\linewidth]{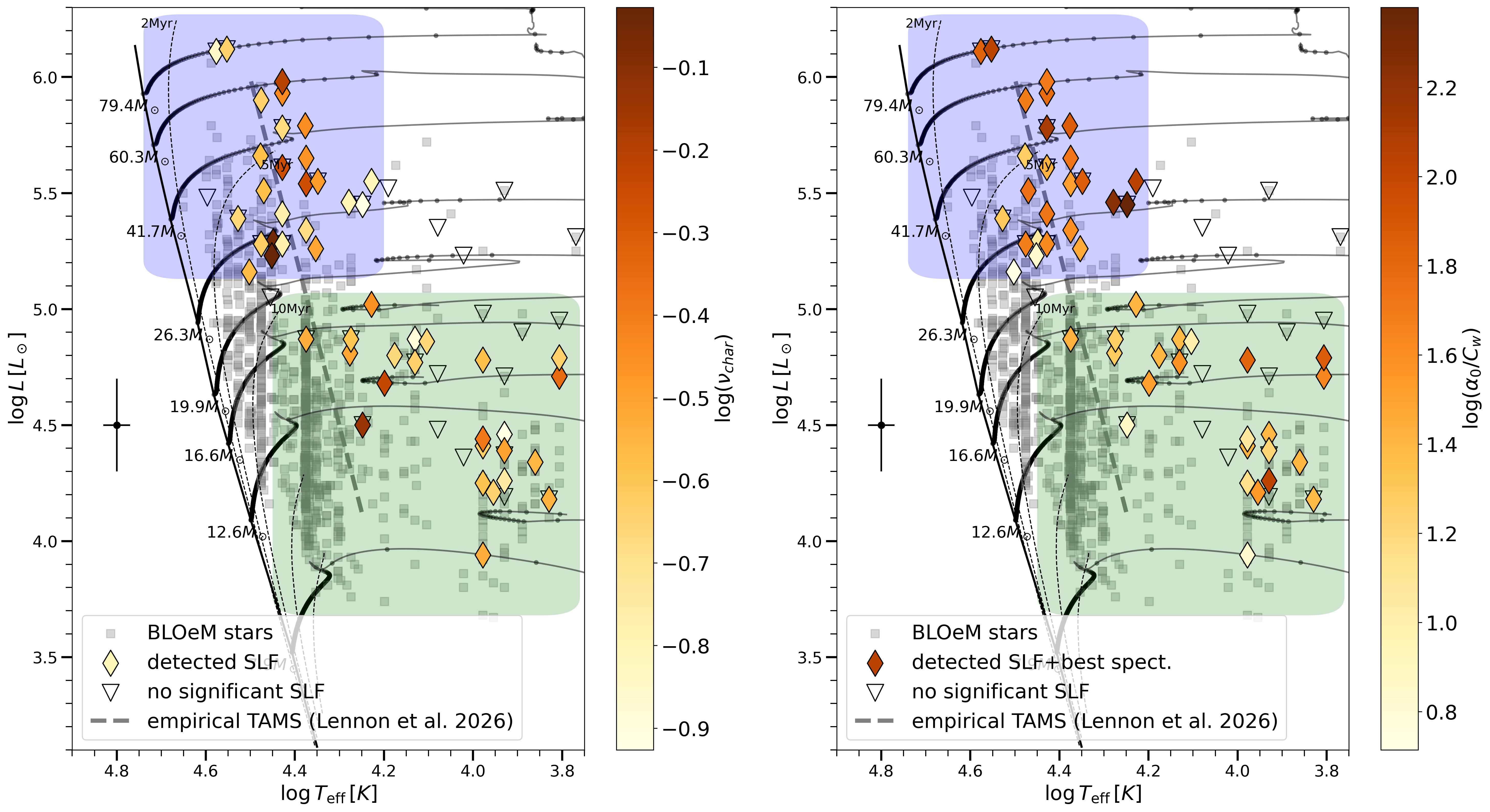}
    \caption{Location in the HR~diagram of the 51 massive stars with significant SLF variability shown as diamonds, which are colour-coded by the characteristic frequency and amplitude relative to the white noise level of their SLF variability in the left and right panels, respectively. When multiple sectors for a star are available, the values shown correspond to the most recent sector available. The grey dashed line is the empirical location of the TAMS based on a decrease in the average $v~{\rm sin}~i$ value for the BLOeM sample from \citep{Lennon+2026}. The sample is further divided in a high-mass main sequence subsample (blue region) and lower mass, mostly post-main sequence stars (green region).}
    \label{fig:HRD SLF}
\end{figure*}

\begin{figure*}
    \centering
    \includegraphics[width=\linewidth]{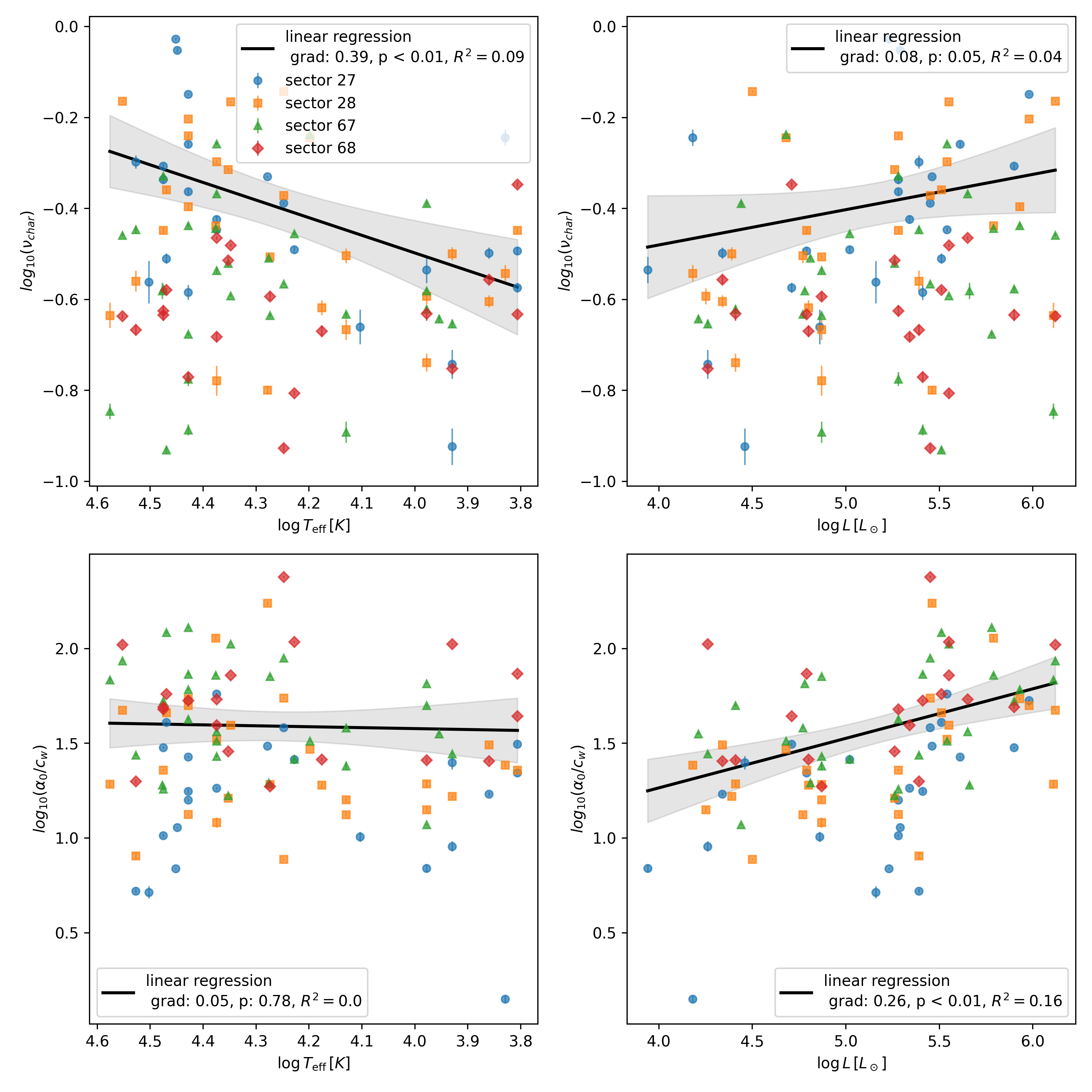}
    \caption{Overview of SLF variability parameters as a function of $T_{\rm eff}$ and luminosity. The different colours indicate the different sectors that have been used to calculate the SLF variability parameters as shown in the legend of the upper left panel. A weighted linear regression combining all sectors is also shown in black, together with a $95\%$ confidence interval. The gradient and $p$-value of these regressions are provided in each panel's legend.}
    \label{fig:SLF_params}
\end{figure*}

Following \citet{BowmanVanDaele2024}, we show the location of all
stars with significant SLF variability in the HR~diagram in Fig.~\ref{fig:HRD SLF}, colour-coded by $\nu_{\rm max}$ in the left panel and colour-coded by the SLF variability amplitude relative to the white noise limit (i.e. $\alpha_0 / C_{\rm w}$) in the right panel. Our work complements the literature in that we find SLF variability is common among different masses and evolutionary stages for SMC massive stars. We also determine the (lack of) correlations between $\nu_{\rm char}$ and $\alpha_0$ with $\log(L)$, $\log(T_{\rm eff})$ using weighted linear regressions, which are shown in Fig.~\ref{fig:SLF_params}, in which multiple sectors per star are utilised. The weights on the data points are determined by the uncertainties on the fitted SLF variability parameters $\nu_{\rm char}$ and $\alpha_0$. Significance of each pair-wise correlation is judged used $p$-values in which the null hypothesis is that there is no correlation between two parameters.

The regression statistics indicate a significant anti-correlation between $\nu_{\rm char}$ and $T_{\rm eff}$, as well as a strong significant correlation between luminosity and $\alpha_0/C_{\rm w}$. In other words, more luminous stars have stronger SLF variability, and cooler (i.e. more evolved) stars have longer period SLF variability. However, we find no statistically significant correlation between $\alpha_0/C_{\rm w}$ and $T_{\rm eff}$. These trends for SMC stars are largely in agreement with what was found for Galactic massive stars by \citet{Bowman+2020aa}. On the other hand, \citet{Bowman+2020aa} found an anti-correlation between $\alpha_0$ and $T_{\rm eff}$, but this is not significant in our sample. We surmise this is due to the lack of stars close to the ZAMS in our sample, since the vast majority of all BLOeM stars with PSF light curves showing SLF variability are near or beyond the TAMS (see Fig.~\ref{fig:HRD SLF}). 
We emphasise that the $R^2$ values of all linear regressions shown in Fig.~\ref{fig:SLF_params} are all $< 0.2$, which implies that the location in the HR~diagram alone is not sufficient to predict a star's SLF variability profile. This is because the sector-to-sector scatter in the SLF variability morphology parameters is quite large, which adds significant scatter in all regressions. Furthermore, what is also not captured in such a small sample of stars is the impact of binarity, differing rotation rates, or magnetic fields, which could all influence the SLF variability parameters, in addition to location in the HR~diagram. Nevertheless, despite this large diversity present, the overall similarities and SLF strength and morphology in this SMC sample and the SMC, LMC and Galactic samples of massive stars from \citet{Bowman+2020aa} and \citet{BowmanVanDaele2024} suggest that the underlying excitation mechanism of SLF variability is largely independent of a star's metallicity.

\begin{figure*}
    \centering
    \includegraphics[width=\linewidth]{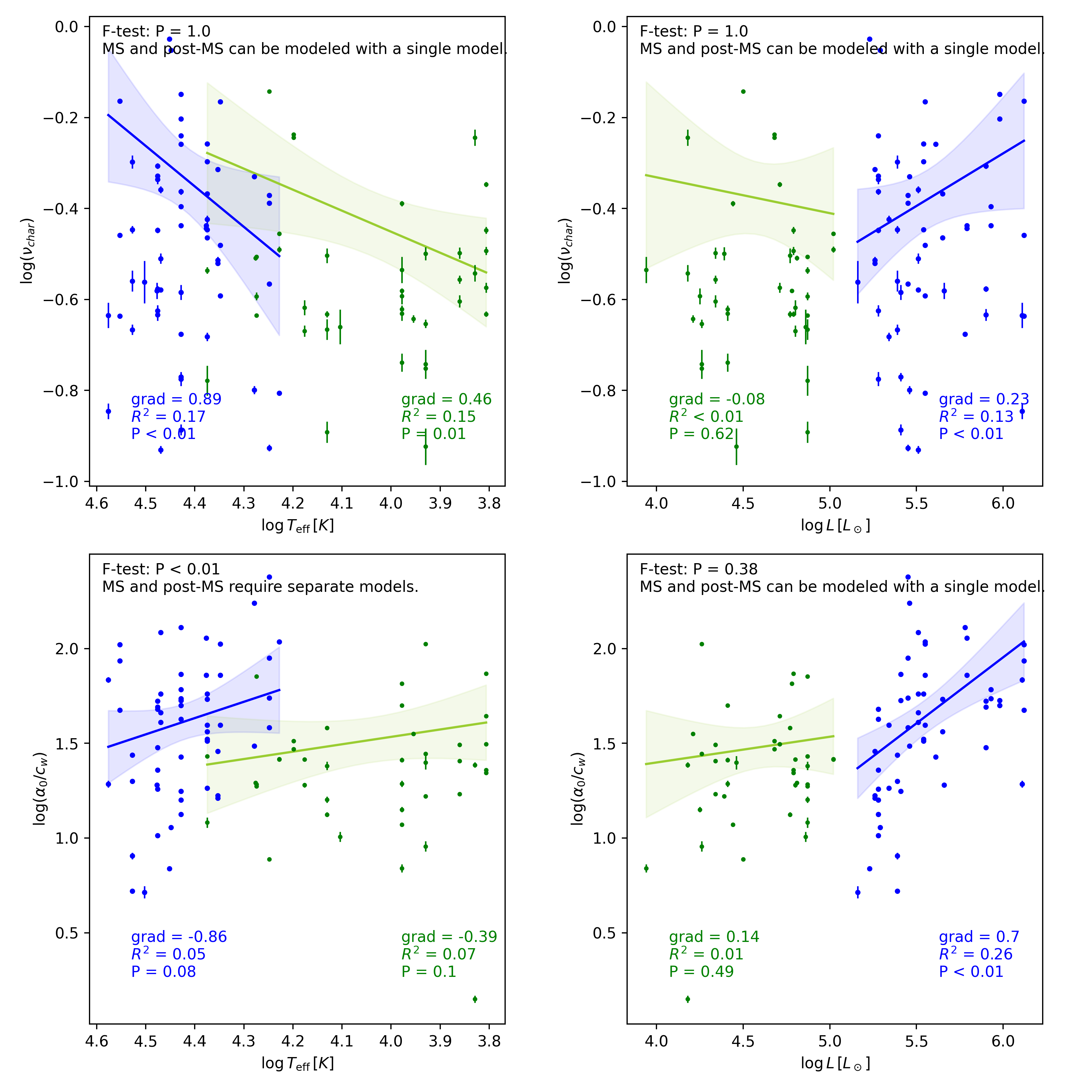}
    \caption{Overview of SLF variability parameters for effective temperature $T_{\rm eff}$ versus luminosity $L$, which were done independently for the high-mass main sequence (blue) and lower-mass post-main sequence (green) subgroups, together with a $95\%$ confidence interval for each linear regression. These two subgroups are shown as regions in the HR~diagram in Fig.~\ref{fig:HRD SLF}. The gradients, $R^2$ and $p$-values of each linear regression are shown for each subgroup in their respective colours. The top textbox shows results of the F‑test for comparing nested linear regression models. Stars in the blue region indicate a luminosity regime in which both main sequence and post-main sequence stars are present, and the systematically higher SLF variability amplitudes are clear for the post-main sequence subgroup of stars.}
    \label{fig:MS_vs_pMS}
\end{figure*}

To investigate whether there is a difference in SLF variability morphology across different evolutionary stages, we divided our sample in two subgroups based on their luminosities: (i) higher luminosity stars (i.e. $\log(L/{\rm L_{\odot}}) > 5.1$) which is also typically high-mass (i.e. $M \gtrsim 20~{\rm M}_{\odot}$) and mostly main-sequence stars as indicated by the blue shaded region in Fig.~\ref{fig:HRD SLF}; and (ii) lower-luminosity stars mainly consisting of post-main sequence lower-mass stars $(8 < M \lesssim 20~{\rm M}_{\odot})$, as indicated by the green region in Fig.~\ref{fig:HRD SLF}. Both subsamples underwent independent linear regressions and an F‑test for comparing nested linear regression models (see e.g. \citealt{book_stats}) was performed, with the results shown in Fig.~\ref{fig:MS_vs_pMS}.
The most apparent difference between the two subsamples is that, on average, the higher luminosity stars have higher SLF variability amplitudes, which was also seen for Galactic massive stars by \citet{Bowman+2020aa}. Interestingly, an important new result from our work is that the F-test supports the conclusion that separate models are required to treat each subsample when SLF variability amplitudes are plotted against $\log(L)$. Additionally, we conclude that $\log(\nu_{\rm char})$ transitions relatively linearly as stars evolve beyond the TAMS in the Hertzsprung gap for SMC massive stars. 

The discrepancy of SLF variability amplitudes in both subsamples can be interpreted considering the mass difference between the two subsamples rather then their different evolutionary stages. Higher mass stars have larger luminosities  thus can allow for higher amplitude SLF variability. This also explains the relatively steep gradient between $\log(\alpha_0/C_w)$ versus $\log(L)$ shown in Fig.~\ref{fig:MS_vs_pMS} for the high-mass subsample. This is because in this region of the HR~diagram, evolutionary tracks are mostly horizontal meaning that luminosity serves as a proxy for stellar mass. 
Of course, expanding both subsamples to include lower mass main-sequence stars (i.e. B dwarfs) and evolved high-mass stars would help further in breaking degeneracies between mass and age. However, high-mass post-main sequence stars, are rare due to their fast evolving nature. This is especially true for stars with birth masses of $M \gtrsim 40$~M$_{\odot}$ because they are subject to the Humphreys-Davidson limit (see \citealt{Humphreys+1979}). 
The lack of main-sequence B dwarfs in our sample corresponds with the limiting brightness we can observe in the SMC with TESS. Nonetheless, such trends would also prove interesting for Galactic massive stars. 

To achieve a detailed understanding of the physical properties of SLF variability, numerical predictions of the near-surface excitation of IGWs need to converge. At present, a significant impact from sub-surface convection zones relies on predictions from 1D stellar evolution models  \citep{Cantiello+2009,Grassitelli+2015,Jermyn+2022} and some multi dimensional numerical simulations \citep{Schultz2022a,Debnath+2024}, but the parameter regime within the HR~diagram in which this would dominate the observable signal is still debated. At the same time, a contribution from IGWs excited by convection in the deep interior to the observable SLF variability is also proposed based on multi-dimensional numerical simulations \citep{Rogers2015, Edelmann+2019ApJ, Thompson+2024_3Dsim, Pathak+2025}. This motivates further study and continued comparison to observations.

A comprehensive picture will also need to include aspects of binary star evolution. The high observed binary fraction of unevolved massive stars (e.g., \citealt{Sana+2012, MarchantBodensteiner2024}) implies that potentially a large number of our sample were involved in a previous binary interaction. This may include mass donors \citep{Pauli+2022}, accretors \citep{Renzo+2021,Richards+2025} and mergers products \citep{Henneco+2024a, Henneco+2024b, Menon+2024}, which both may appear as OB dwarfs, but with atypical luminosity-to-mass ratios compared to single stars. This may lead to scatter of the SLF properties in the HR~diagram. Once accurate gravities are determined for our stars, a re-analysis based on their position in the spectroscopic HR diagram would therefore be promising. We believe that progress on the methodology and observational strategy reported here indicates that a profound empirical discrimination of the theories and models which predict SLF variability is soon within our reach. Regardless, our sample is sufficient to illustrate the common yet diverse morphology of SLF variability within SMC massive stars. Our data shows clear indications that the morphology of SLF variability correlates with different physical properties of the stars, such as stellar mass. However, there is a lot of room for further investigation exploring the role of rotation, binarity, and chemical mixing, and how these contribute to scatter in the inferred parameters, which will be possible once a bigger sample is available.

\section{Conclusions}
\label{sec: conclusions}

In this work, we have developed a new software tool, {\sc Lemons}, to extract light curves by modelling PSFs of a sample of massive star targets and minimise contamination from nearby sources in TESS FFI images. Our PSF fitting software tool is optimised for massive stars, which are often variable with a large range of periods and amplitudes, hence require flexibility to extract optimal light curves. For example, we demonstrate that PSF centroid and broadness as free parameters yield robust light curves. Moreover, we have illustrated that this method mitigates contamination issues, compared to the standard SAP approach, in particular for faint ($13 \lesssim G \lesssim15$~mag) targets and stars in crowded regions for a subset of the BLOeM massive stars in the SMC.

We provide a new sample of 91 massive SMC stars with TESS light curves, along with a variability classification. In these low-metallicity stars, we observe different types of variability, including coherent pulsators, binarity, rotational modulation, and SLF variability. Our light curves also show that SLF variability is common in massive stars in the SMC and that it has a similar morphology to massive stars in the LMC and Milky Way galaxies. The similar morphologies of SLF variability in both main-sequence and post-main-sequence massive stars in the SMC studied in this work, the comparable SLF variability for massive stars in the SMC, LMC and the Galaxy studied by \citet{Bowman+2020aa} and \citet{BowmanVanDaele2024}, combined with the expectation that sub-surface convection is dependent on metallicity \citep{Jermyn+2022}, suggests there is still much theoretical work to do in understanding the excitation mechanism of IGWs.

When combined with the ongoing BLOeM spectroscopic survey of the SMC, these PSF light curves allow ensemble studies for binarity and pulsations in low-metallicity massive stars. Asteroseismology of such a unique sample would open a new window for probing stellar interiors at low metallicities and thus improve stellar evolution models. Finally, our study serves as a proof-of-concept for demonstrating the need to analyse pixel data (i.e. `imagettes') of faint massive stars with the upcoming ESA PLATO mission \citep{Rauer2025}, which is scheduled to launch in early-2027 and will observe the LMC within its 2-yr pointing in the southern hemisphere. Moreover, the proposed ESA HAYDN mission \citep{Miglio+2021} will target young, dense environments including many pulsating massive stars, for example, the young open cluster h \& $\chi$~Per \citep{Saesen+2013, Nardini+2025}, and our PSF software tool adds validity for future scientific results in this regard. Specifically, to most effectively analyse faint stars in crowded regions such as the LMC and SMC, image pixel data are critical for reasons of contamination. With excellent CCD pixel image data provided by PLATO, we will be able to apply our PSF fitting software and extract the light curves of thousands of massive stars for the wider community.

\section*{Acknowledgements}

The TESS data presented in this paper were obtained from the Mikulski Archive for Space Telescopes (MAST) at the Space Telescope Science Institute (STScI), which is operated by the Association of Universities for Research in Astronomy, Inc., under NASA contract NAS5-26555. Support to MAST for these data is provided by the NASA Office of Space Science via grant NAG5-7584 and by other grants and contracts. Funding for the TESS mission is provided by the NASA Explorer Program.

The authors acknowledge M. Vanrespaille and M. Cantiello for comments on a draft of this manuscript, and the referee for their constructive feedback.

The authors gratefully acknowledge UK Research and Innovation (UKRI) in the form of a Frontier Research grant under the UK government's ERC Horizon Europe funding guarantee (SYMPHONY; PI Bowman; grant number: EP/Y031059/1), and a Royal Society University Research Fellowship (PI Bowman; grant number: URF{\textbackslash}R1{\textbackslash}231631). TS acknowledges support from the European Research Council (ERC) under the European Union's Horizon 2020 research and innovation program (grant agreement 101164755/METAL) and from the Israel Science Foundation (ISF) under grant number 0603225041. DP acknowledges financial support from the FWO in the form of a junior postdoctoral fellowship No.\,1256225N.

\section*{Data Availability}

The TESS data presented in this paper were obtained from the Mikulski Archive for Space Telescopes (MAST; \url{https://archive.stsci.edu/missions-and-data/tess}). This research made use of 
{\tt astropy} \citep{astropy}, 
{\tt matplotlib} \citep{matplotlib}, 
{\tt numpy} \citep{numpy}, 
{\tt Pandas} \citep{pandas}, 
{\tt Lightkurve} \citep{Lightkurve,2018_lightkurve}, 
{\tt TESScut} \citep{TESScut_ref}, 
{\tt Photutils} \citep{photutils} and 
{\tt Period04} \citep{period04_2005}.

The {\sc Lemons} software package is open source and accessible via \url{github.com/pieterjanv314/Lemons}. For the purpose of open access, the authors have applied a CC BY licence to the author accepted manuscript version: \url{https://https://arxiv.org/abs/2605.15757}.
The data underlying this article are available in a Zenodo repository, at \url{https://zenodo.org/records/20540863}.



\bibliographystyle{mnras}
\bibliography{master_bib}


\newpage
\onecolumn

\appendix
\section{Variability classification for all PSF observed BLOeM stars}

    \captionsetup{width=\textwidth}
    \captionof{table}{Overview of SMC stars with PSF light curves, with spectral types from \citet{BLOeM2024}} and the dominant TESS photometric variability type identified in this work. Light curves without significant variability or with instrument dominated signal are left blank. Categories are SLF (stochastic low frequency) variability, EV (ellipsoidal variability), RotMod (rotational modulation) and pulsator to denote coherent pulsation modes.
    \label{tab:psf-lc}
    \begin{longtable}{ccccc}
      \toprule
      BLOeM ID & GAIA DR3 ID & Tmag & Spectral Type & TESS PSF light curve variability \\
      \midrule
      \endfirsthead
      \multicolumn{5}{c}{\tablename\ \thetable\ (continued)}\\
      \toprule
      BLOeM ID & GAIA DR3 & Tmag & Spectral Type & TESS PSF light curve variability \\
      \midrule
      \endhead
      \midrule
      \multicolumn{5}{r}{\textit{Continued on next page}}\\
      \endfoot
      \bottomrule
      \endlastfoot
1-009 & 4690520387983090432 & 12.6 & B1 Ia & SLF\\ 
1-011 & 4690506712804881792 & 14.3 & B1.5 II & EV or RotMod\\ 
1-016 & 4690506914646214912 & 12.1 & A2 Ib & \\ 
1-028 & 4690519769509887488 & 12.3 & B8 Iab/Ia & \\ 
1-031 & 4690507670561677696 & 12.6 & A0 Iab & \\ 
1-039 & 4690503306873686912 & 12.8 & A2 II/Ib & SLF\\ 
1-040 & 4690525438865707904 & 13.7 & O9.7 III:n e & pulsator\\ 
1-051 & 4690502280398446720 & 10.8 & A0 Ia & pulsator\\ 
1-053 & 4690502280398443776 & 13.9 & B1 Ib & \\ 
1-062 & 4690507881037706112 & 13.1 & B8 Iab & \\ 
1-065 & 4687505389671548672 & 12.9 & A7 Ib & \\ 
1-085 & 4687503946562429184 & 13.1 & A7 Ib/ab & \\ 
1-095 & 4690508632639753984 & 12.9 & B1 Ia & SLF \\ 
1-106 & 4687505801988031232 & 13.6 & B1.5 e & SLF\\ 
1-111 & 4687507584364583040 & 11.4 & B3 Ia & SLF \\ 
1-112 & 4687507245097014144 & 11.2 & B9 Ia & \\ 
2-007 & 4688978838366817280 & 12.7 & O9.5 II-I & SLF\\ 
2-054 & 4688966292708802304 & 12.1 & A0 Ia & SLF\\ 
2-065 & 4688967568356557440 & 10.7 & F5: &\\ 
2-068 & 4688980414561766016 & 13.0 & B9 Iab & SLF\\ 
2-072 & 4685959373284393728 & 13.9 & A2 II/Ib & \\ 
2-080 & 4688962517474878720 & 13.5 & A2 II/Ib & \\ 
2-092 & 4688967289176997504 & 12.3 & B8 Iab & \\ 
2-093 & 4688967362197739392 & 10.8 & B8 Ia & pulsator\\ 
2-113 & 4685983420730442624 & 12.4 & B2.5 Ia & SLF\\ 
2-116 & 4688986534938381184 & 12.9 & sgB[e] & SLF\\ 
3-012 & 4685854571724688768 & 12.9 & sgB[e] & SLF\\ 
3-037 & 4685835884339219456 & 12.3 & B3 Ia & SLF\\ 
3-042 & 4685947553531340800 & 11.2 & O6 I(f) & EV or RotMod\\ 
3-043 & 4685943121123943680 & 11.0 & A2 Iab & SLF\\ 
3-061 & 4685943254211136000 & 12.0 & B3 Ib & SLF\\ 
3-106 & 4685945010910000640 & 13.6 & B5 Ib & \\ 
4-001 & 4689018661298288000 & 13.3 & A1 Ib & SLF\\ 
4-006 & 4689002851458876800 & 11.4 & F2: & \\ 
4-020 & 4689003027613801216 & 12.9 & B1 Iab-Ib & SLF\\ 
4-055 & 4690522273449262592 & 12.5 & sgB[e] & SLF\\ 
4-058 & 4690516677131714432 & 11.2 & O7 Iaf$^{+}$ & SLF\\ 
4-072 & 4690500184455105536 & 12.5 & B9 Ia & \\ 
4-078 & 4690503826587160832 & 12.3 & B1 Ia & SLF\\ 
4-079 & 4690521762372716160 & 13.3 & A0 Ib & \\ 
4-084 & 4690522002890799488 & 12.8 & A0 Iab & SLF\\ 
4-103 & 4690500047016009728 & 12.6 & A5 Ib & \\ 
5-005 & 4687504182822540416 & 13.4 & F2: & pulsator\\ 
5-012 & 4687499720314663296 & 13.5 & A2 II/Ib & \\ 
5-014 & 4687487900566335744 & 13.0 & A0 Ib & SLF\\ 
5-029 & 4687487213367243392 & 12.9 & A7 Ib & pulsator\\ 
5-050 & 4687501953697501568 & 14.0 & O9.7 V: & EV or RotMod\\ 
5-067 & 4687485387976612736 & 12.4 & A0 Iab & \\ 
5-071 & 4687437254305162880 & 13.5 & O8.5: Ib & SLF\\ 
5-077 & 4687485491055825536 & 12.4 & B2.5 Ia & SLF\\ 
5-090 & 4687513910877142528 & 14.1 & O9.5 III & \\ 
5-091 & 4687437426103845376 & 13.1 & B8 Ib & \\ 
5-102 & 4687508413319210368 & 13.4 & F0: & \\ 
5-104 & 4687512948804514176 & 13.8 & B2.5 Ia & SLF\\ 
5-105 & 4687509749039520768 & 13.4 & B0.7 II & SLF\\ 
6-001 & 4687162582543422080 & 14.0 & B0.5 III: e & EV or RotMod\\ 
6-002 & 4687161929708394752 & 12.8 & A2 Ib & \\ 
6-005 & 4687161895348646400 & 13.8 & O9.7 II-Ib(n) & SLF \\ 
6-006 & 4687158592506675328 & 12.3 & F2: & SLF\\ 
6-007 & 4687157286836619520 & 15.2 & B5 II & \\ 
6-008 & 4687159868124113152 & 10.0 & A2 Ia & \\ 
6-015 & 4686408111398825728 & 12.6 & F2: & \\ 
6-017 & 4687165125163949952 & 13.4 & B1 Ib & SLF\\ 
6-024 & 4687164055705068288 & 11.5 & A5 Iab & \\ 
6-052 & 4687165709279578368 & 12.8 & A2 II/Ib & SLF \\ 
6-067 & 4686413230999834496 & 14.3 & O9.7 III & SLF\\ 
6-072 & 4686413540237697408 & 14.4 & B3 II & \\ 
6-076 & 4686414467950355072 & 13.0 & B0 Ib & SLF\\ 
6-080 & 4686413437158238208 & 12.1 & B0 Ia & SLF\\ 
6-084 & 4687167152388524032 & 14.2 & O9.7 I:(n) & SLF\\ 
6-093 & 4686413810818254464 & 14.2 & O8.5 V & \\ 
6-095 & 4686410550940229888 & 14.0 & A0 Ib & SLF\\ 
6-105 & 4686410997616777984 & 14.3 & O6 V:n & EV or RotMod\\ 
6-108 & 4686410688379157888 & 13.2 & A0 Iab & SLF\\ 
6-116 & 4686417148009897344 & 13.4 & A7 Iab & \\ 
7-004 & 4685987857514223744 & 13.7 & A0 Iab & \\ 
7-064 & 4685972636146149504 & 12.3 & B0 Ia & SLF\\ 
7-075 & 4685976621875402368 & 13.0 & B8 Ib & SLF\\ 
7-091 & 4687479173187523584 & 11.3 & A5 Iab & \\ 
7-108 & 4687426087389568768 & 14.0 & B2.5 Ia & SLF\\ 
7-111 & 4687478138084449536 & 12.6 & B0 Ia & SLF\\ 
8-001 & 4689054670286883968 & 13.0 & B8 Ib & \\ 
8-006 & 4689037181200454656 & 13.8 & B2.5 II-Ib & EV or RotMod\\ 
8-008 & 4689054945164735616 & 12.8 & B1 Iab & SLF\\ 
8-010 & 4689033539046230400 & 11.4 & A0 Ia & pulsator\\ 
8-022 & 4689062298150064640 & 13.8 & B0.5 II & EV or RotMod\\ 
8-053 & 4689074598938187776 & 14.3 & O9 III & \\ 
8-072 & 4689057346006300544 & 11.9 & F2: &\\ 
8-082 & 4689061920192649984 & 13.1 & B9 Iab & \\ 
8-097 & 4689058999614762368 & 13.5 & F0: & \\ 
8-116 & 4689058892200836096 & 13.0 & A2 Ib & \\ 
    \end{longtable}
\twocolumn

\section{Periodograms of TESS light curves}

\begin{figure*}
  \centering
  \includegraphics[width=\linewidth]{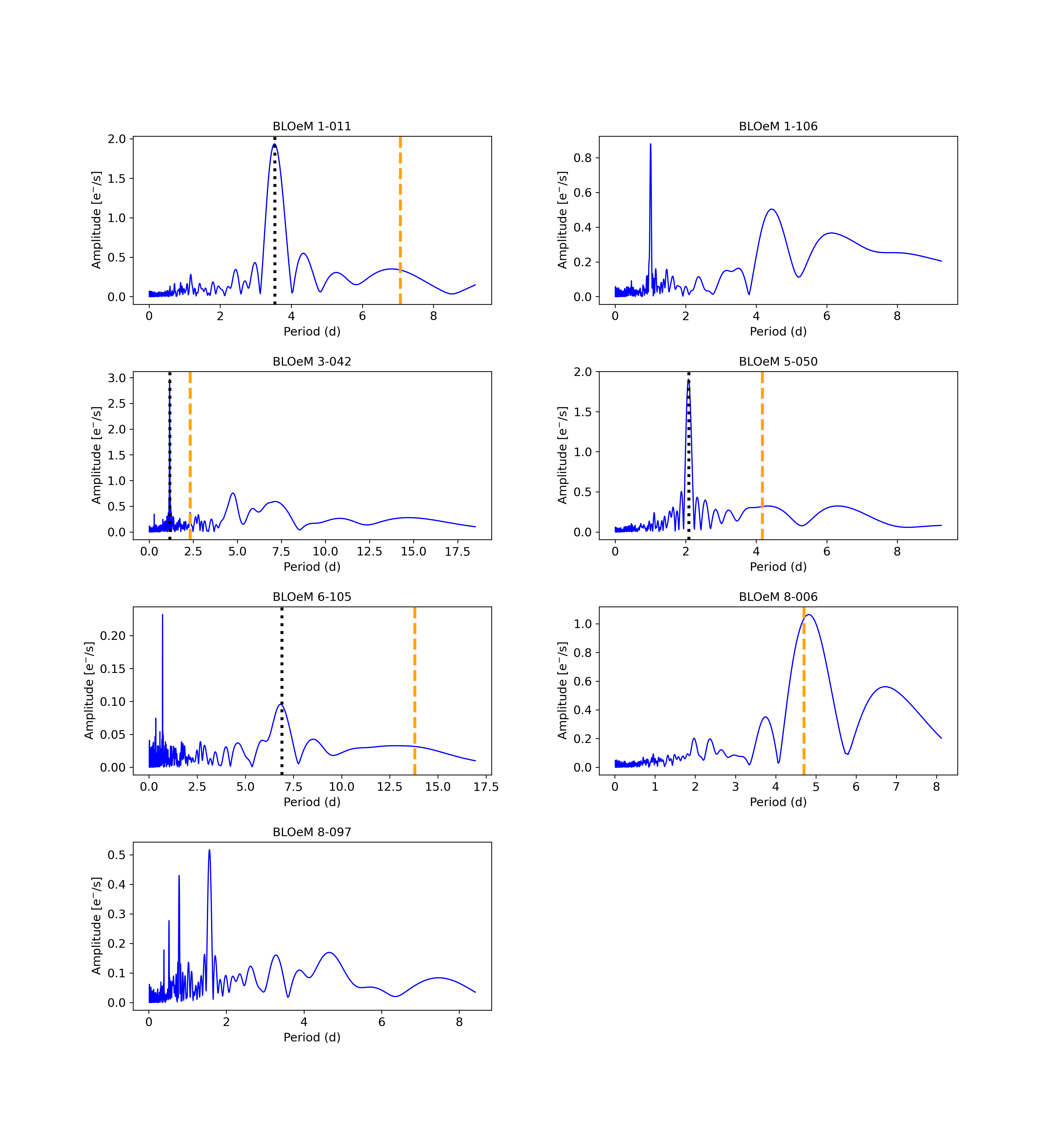}
  \caption{Periodograms of PSF TESS light curves of confirmed and candidate binary systems, which are presented in Fig.~\ref{fig: bin_spec} and Fig.~\ref{fig: bin_nospec}, respectively. Spectroscopic BLOeM periods are shown with an orange dashed vertical line. Half the BLOeM period is shown with a black-dotted line when it matches the TESS variability. All these periods are quantified in Table~\ref{table:binaries}.}
  \label{fig: appendix: binaries}
\end{figure*}

The Lomb-Scargle periodograms of our PSF TESS light curves for the confirmed and candidate binaries are shown in Fig.~\ref{fig: appendix: binaries}.

\bsp	
\label{lastpage}
\end{document}